\documentclass[
  prl,
  twocolumn,
  superscriptaddress,
  longbibliography,
  nofootinbib,
  nobibnotes
]{revtex4-2}

\DeclareUnicodeCharacter{03F5}{\ensuremath{\epsilon}}
\usepackage{amsmath,amssymb,bm}
\usepackage{xcolor}
\usepackage{graphicx,dcolumn,braket,slashed}
\usepackage[colorlinks=true,linkcolor=blue,citecolor=blue,urlcolor=blue]{hyperref}
\usepackage{bookmark}

\newcommand{\Ree}{\operatorname{Re}}
\newcommand{\Imm}{\operatorname{Im}}

\newcommand{\avgT}[1]{\langle #1 \rangle_T}

\usepackage[normalem]{ulem}
\usepackage{placeins}

\begin{document}

\title{Energy-Centroid Trembling near Exceptional Points in Photonic Time Crystals}

\author{Kyungmin Lee}
\thanks{These authors contributed equally to this work.}
\affiliation{Department of Physics, Korea Advanced Institute of Science and Technology, Daejeon 34141, Republic of Korea}

\author{Younsung Kim}
\thanks{These authors contributed equally to this work.}
\affiliation{Department of Physics, Korea Advanced Institute of Science and Technology, Daejeon 34141, Republic of Korea}

\author{Minwook Kyung}
\affiliation{Department of Physics, Korea Advanced Institute of Science and Technology, Daejeon 34141, Republic of Korea}

\author{Seungmin Yeon}
\affiliation{Department of Physics, Korea Advanced Institute of Science and Technology, Daejeon 34141, Republic of Korea}

\author{Bumki Min}
\email{bmin@kaist.ac.kr}
\affiliation{Department of Physics, Korea Advanced Institute of Science and Technology, Daejeon 34141, Republic of Korea}

\date{\today}

\begin{abstract}
In a photonic time crystal with a modulated Lorentz restoring coefficient, modulation reweights the potential energy of the polarization and loss drains its kinetic energy; when these sectors are offset within a pulse, the energy centroid trembles in every cycle while the Poynting-flux velocity stays bounded.
Near the gap-edge exceptional point the mode geometry amplifies these offsets, and for a narrow-band pulse on one passband Floquet branch the trembling-velocity amplitude generically grows as the inverse square root of the momentum detuning, rounded by finite bandwidth.
Loss modulated in phase or in antiphase with the restoring coefficient can change the trembling coefficient while the band-slope and Petermann coefficients barely move.
In the lossy example studied, a forward field pulse launched with the polarization at rest exhibits delayed bursts near the beat recurrence after substantial energy loss, with monochromatic amplitudes following the inverse three-halves power of the detuning over the computed range.
Intracycle energy exchange and preparation thus shape the centroid motion beyond what the band slope and the Floquet Petermann factor determine.
\end{abstract}

\maketitle

\emph{Introduction.---}At a regular Bragg band edge of a lossless one-dimensional photonic crystal, counterpropagating waves form a standing wave, and both the group velocity and the energy-transport velocity vanish.
Temporal periodicity produces a contrasting band edge.
In a spatially uniform photonic time crystal (PTC)~\cite{galiffi2022photonics,lustig2023photonic,asgari2024theory,kim2026quantum}, modulation conserves the wavenumber and couples positive- and negative-frequency components near parametric resonance.
The resulting Floquet modes coalesce at exceptional points (EPs) bounding a momentum gap~\cite{zurita2009reflection,reyes2015observation,wang2018photonic,park2022revealing,lyubarov2022amplified,wang2023metasurface,wang2025expanding,bae2025quantum,xiong2025observation,guo2026plasmonic}.
Approaching an EP from the passband, the band slope diverges~\cite{martinez2017standing}, yet the Poynting-flux velocity remains bounded by the background light speed in the Maxwell--Lorentz medium considered here~\cite{LEE2026100636}.
How does this singular mode structure manifest itself in the motion of stored energy within each modulation cycle?

\begin{figure}[t]
\centering
\includegraphics[width=\columnwidth]{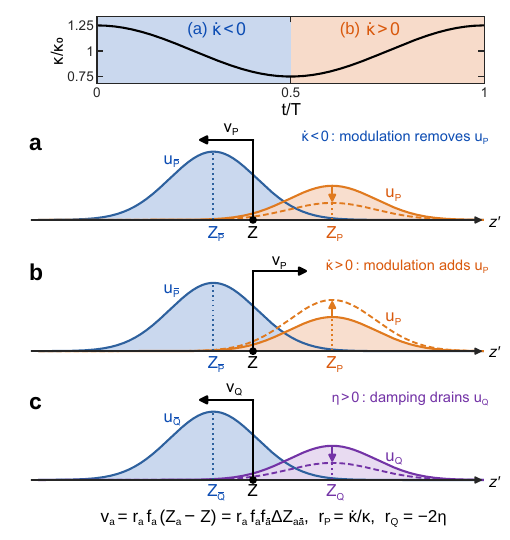}
\caption{\label{fig:Fig1}
Reweighting an offset sector moves the energy centroid.
Solid and dashed profiles show a sector ahead of its complement before and after reweighting by the indicated source term alone.
Dotted lines mark the sector centroids and black markers the total centroid $Z$.
Arrows give the contribution $v_a=r_a f_a(Z_a-Z)=r_a f_a f_{\bar a}\Delta Z_{a\bar a}$ of each channel to the centroid velocity, with $r_P=\dot\kappa/\kappa$ and $r_Q=-2\eta$, in the frame $z'=z-\int v_{\rm flux}\,\mathrm{d}t$.
(a) Softening removes and (b) stiffening adds potential energy in the $P|\bar P$ partition, and (c) damping drains kinetic energy at all phases in the $Q|\bar Q$ partition.
Top: $\kappa(t)/\kappa_0$ over one period.
A sector behind $Z$ reverses each arrow.
Profiles are illustrative.}
\end{figure}

A dispersive medium stores energy in its polarization as well as in the field, and in a Maxwell--Lorentz description the stored energy separates into the electromagnetic energy, the potential energy of the bound-charge displacement and the kinetic energy of the charge motion~\cite{loudon1970propagation,loudon1997propagation,sloan2024optical,sun2025formulation,lee2026analogs,allard2026broadband,sustaeta2026fluctuational}.
For a PTC with time-periodic permittivity, the cycle-averaged energy velocity has been shown to stay bounded and the divergent band slope has been traced to a modulation-driven geometric drift set by mismatched electric and magnetic phase connections, and the energy balance and the velocity bound have been extended to a Lorentz resonance with a modulated restoring force~\cite{LEE2026100636}.
In non-Hermitian coupled-resonator waveguides, wave-packet group velocities can differ from velocities inferred from the probability flux~\cite{PhysRevA.90.053819}.
In static non-Hermitian bands, Berry-connection terms correct the weight and the velocity of a driven packet~\cite{silberstein2020berry}, and near an EP a divergent quantum metric produces anomalous acceleration and drift~\cite{solnyshkov2021quantum}.
In a PTC, however, modulation of the restoring coefficient exchanges energy with the potential-energy sector of the polarization within every period, while damping drains its kinetic-energy sector, so the sector weights of the energy centroid change inside each cycle.

We resolve the resulting centroid motion near the Floquet EP with an exact identity that splits the centroid velocity into a Poynting-flux part bounded by the background light speed and a source part, set by how fast modulation and loss exchange energy with each sector and by how far that sector sits from the rest of the pulse.
The picture is familiar from pulse amplification, where gain on the leading part of an optical pulse pulls its energy center forward by raising the weight of that part.
For a narrow-band pulse on one Floquet branch, the period average of the source part carries the divergent band slope, as in the cycle-averaged picture~\cite{LEE2026100636}, and its zero-mean part, which the cycle average removes, makes the energy centroid tremble within each modulation period.
This source-driven trembling recalls Zitterbewegung, the coherent oscillation of a wave-packet centroid produced by interference between positive- and negative-energy components~\cite{schrodinger1930kraftefreie,longhi2010photonic,dreisow2010classical,gerritsma2010quantum,leblanc2013direct,lovett2023observation,hu2025zitterbewegung,yang2026non}.
Near the gap, the dominant interfering harmonics of this packet are inherited from the positive- and negative-frequency static branches, while every stored-energy sector stays nonnegative.
Modulation and loss turn the resulting sector offsets into motion.
As the EP is approached, the trembling velocity grows as the inverse square root of the momentum detuning, while the weighted squared sector separations and the energy-norm Petermann factor grow as its inverse.
The three laws follow from one Jordan chain, yet loss modulation in phase or in antiphase with the restoring coefficient can change the trembling coefficient while leaving the band-slope and Petermann coefficients essentially unchanged.
A finite spectral width rounds the growth into a finite response.
The preparation then sets the later motion: in our eigenvector gauge the trembling of an eigen-packet with a real envelope fades once modes inside the gap dominate the pulse, a quadratic spectral chirp keeps a finite gap-side response over the computed times, and a pulse launched in the fields alone excites both coalescing modes and trembles in bursts near their beat recurrence.

\emph{Energy sectors and centroid motion.---}We consider a one-dimensional Maxwell--Lorentz medium whose Lorentz oscillator has a restoring coefficient modulated in time.
Keeping the polarization $P$ and its velocity $Q\equiv\dot P$ as dynamical variables resolves the material energy into a potential and a kinetic sector.
With background permittivity $\epsilon_b$ and permeability $\mu$, the source-free dynamics read
\begin{equation}
\begin{aligned}
&\partial_z E=-\mu\,\partial_t H,\qquad
\partial_z H=-\partial_t(\epsilon_b E+P),\\
&\ddot P+\eta\dot P+\kappa(t)P=\alpha E,
\end{aligned}
\label{eq:maxwell_lorentz}
\end{equation}
where $\kappa(t)=\kappa_0[1+m\cos(\Omega t)]$ is the restoring coefficient of period $T=2\pi/\Omega$, $\eta\ge0$ is the damping rate and $\alpha>0$ sets the strength of the polarization response to the field.
These are oscillators of fixed density with a prescribed stiffness and a local viscous loss, and in the continuum limit the same equations describe a transmission line shunt-loaded by series resonators with modulated capacitance and resistance~\cite{SM}.
With complex fields averaged over their real and imaginary parts, which for pulses with spectra confined to $k>0$ give the sector energies and centroid of the single real field at every instant~\cite{SM}, the field--matter energy density and its exact local balance read
\begin{equation}
\begin{aligned}
&u_{\rm tot}=\frac14\bigl(\epsilon_b|E|^2+\mu|H|^2\bigr)
 +\frac{1}{4\alpha}\bigl[\kappa(t)|P|^2+|Q|^2\bigr],\\
&\partial_t u_{\rm tot}+\partial_z S_z=\frac{\dot\kappa}{\kappa}u_P-2\eta u_Q\equiv s,
\end{aligned}
\label{eq:energy_balance}
\end{equation}
where the four terms of $u_{\rm tot}$ are the electric, magnetic, polarization-potential and polarization-kinetic sectors $u_E$, $u_H$, $u_P$ and $u_Q$, all nonnegative for positive $\kappa_0$, $\epsilon_b$, $\mu$ and $|m|<1$, and $S_z=\Ree(EH^*)/2$ is the Poynting flux~\cite{SM}.
Modulation adds potential energy at the rate $(\dot\kappa/\kappa)u_P$ while the restoring coefficient rises and removes it while it falls, and damping drains kinetic energy at the rate $2\eta u_Q$.

We define the total energy $U=\int u_{\rm tot}\,\mathrm{d}z$, its centroid $Z=U^{-1}\int z\,u_{\rm tot}\,\mathrm{d}z$ and, for each nonempty sector $a$ with complement $\bar a$, the sector energy $U_a=\int u_a\,\mathrm{d}z$, the energy fraction $f_a=U_a/U$ and the sector centroid $Z_a=U_a^{-1}\int z\,u_a\,\mathrm{d}z$.
Integrating Eq.~\eqref{eq:energy_balance} over a localized pulse gives the exact centroid equation~\cite{SM}
\begin{equation}
\begin{aligned}
\dot Z&=v_{\rm flux}+v_{\rm src},\qquad
|v_{\rm flux}|\le c_b=(\epsilon_b\mu)^{-1/2},\\
v_{\rm src}&=\frac{\dot\kappa}{\kappa}f_P f_{\bar P}\Delta Z_{P\bar P}
-2\eta f_Q f_{\bar Q}\Delta Z_{Q\bar Q},
\end{aligned}
\label{eq:centroid_identity}
\end{equation}
where $\Delta Z_{a\bar a}=Z_a-Z_{\bar a}$, $v_{\rm flux}=U^{-1}\int S_z\,\mathrm{d}z$, and the two terms of $v_{\rm src}$ define the modulation channel $v_P$ and the damping channel $v_Q$.
Flux transport stays within $c_b$ because $|S_z|\le c_b(u_E+u_H)$ at every point and the nonnegative material energy only enlarges $U$.
Energy added to a displaced sector pulls the centroid toward that sector and energy removed pushes it away, in proportion to the weighted offset $f_a f_{\bar a}\Delta Z_{a\bar a}$.
The sign of the modulation contribution therefore depends on both the order of the $P$ and $\bar P$ centroids and the sign of $\dot\kappa/\kappa$.
For example, when the potential-energy sector lies ahead of its complement, softening gives a backward modulation contribution and stiffening gives a forward contribution [Fig.~\ref{fig:Fig1}(a,b)].
Nonzero damping gives a backward source contribution whenever the kinetic-energy sector lies ahead of its complement [Fig.~\ref{fig:Fig1}(c)].

Because the inter-sector part of the energy variance is $f_a f_{\bar a}(\Delta Z_{a\bar a})^2$, a pulse with finite rms width $\sigma_z$ also obeys
\begin{equation}
|v_{\rm src}|\le
\left(\frac12\left|\frac{\dot\kappa}{\kappa}\right|+\eta\right)\sigma_z.
\label{eq:source_width_bound}
\end{equation}
For wide pulses the bound permits source velocities above $c_b$, and those computed below reach several times $c_b$.
They describe how modulation and damping move the weights of stored energy~\cite{LEE2026100636}.
For a single-branch pulse with nonzero spectral weight at the EP, the square-root cusp of the mode leaves the centroids finite but makes the spatial variance $\sigma_z^2$ diverge logarithmically.
Finite bandwidth rounds the trembling-velocity growth (see below); Eq.~\eqref{eq:source_width_bound}, however, also requires finite spatial variance.
With a globally regular spectral completion, smoothing the mode profile over an interval $\varepsilon_k\ll\sigma_k$ about the EP gives a launch with finite $\sigma_z$ and preserves the leading fixed-window response of Eq.~\eqref{eq:finite_bandwidth_scaling} as $\varepsilon_k/\sigma_k\to0$~\cite{SM}.

\begin{figure*}[t]
\centering
\includegraphics[width=\textwidth]{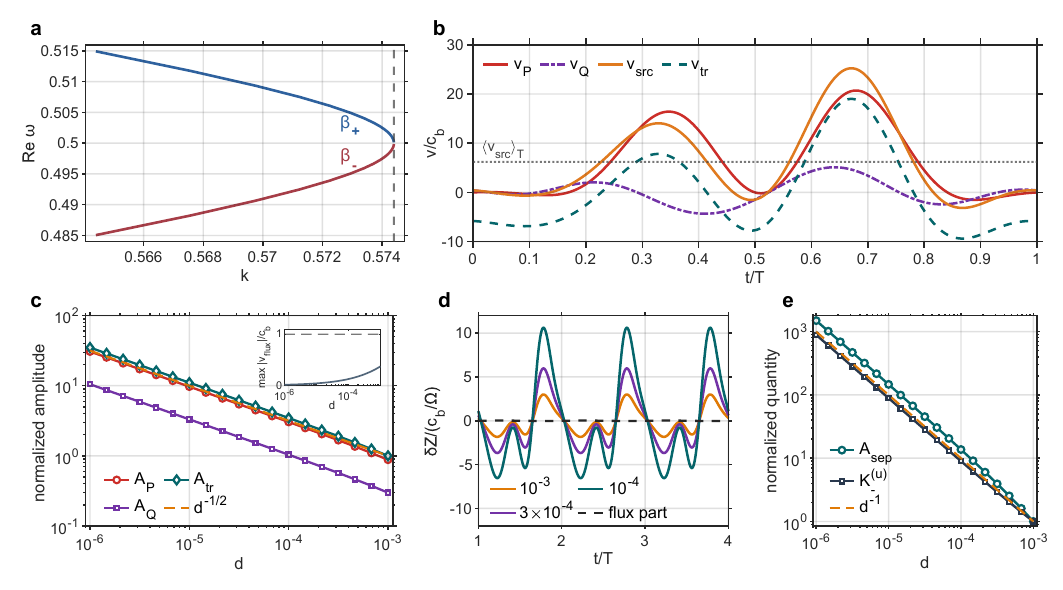}
\caption{\label{fig:Fig2}
Single-branch trembling grows near the exceptional point while the flux velocity stays bounded.
(a) Real quasifrequencies of the two resonant Floquet branches near the lower gap edge; the dashed line marks $k_{\rm EP}$, and $d\equiv k_{\rm EP}-k$.
(b) Modulation channel $v_P$, damping channel $v_Q$, their sum $v_{\rm src}$ and the trembling velocity $v_{\rm tr}$ on the $\beta=-$ branch at $d=10^{-4}$; the dotted line marks $\avgT{v_{\rm src}}$.
(c) Centered rms amplitudes of $v_P$, $v_Q$ and $v_{\rm tr}$, normalized by $A_{\rm tr}$ at $d=10^{-3}$; dashed guide, $d^{-1/2}$.
Inset: $\max_t|v_{\rm flux}|/c_b$.
(d) Mean-centered centroid displacement $\delta Z$ after subtraction of the uniform drift $\bar v t$, with $\bar v$ and the centering taken over $1\le t/T\le4$, for single-branch packets of spectral width $\sigma_k=d_0/20$, where $d_0=k_{\rm EP}-k_0$ is the detuning of the packet center $k_0$; dashed, flux contribution at $d_0=10^{-4}$.
(e) Energy-norm Petermann factor $K_-^{(u)}$ and centered rms $A_{\rm sep}$ of $S_{\rm sep}$, each normalized at $d=10^{-3}$; dashed guide, $d^{-1}$.
$A_{\rm tr}$ is the centered rms of the source velocity $v_{\rm src}$ over one period.
Parameters: $\epsilon_b=\mu=\Omega=1$, $\alpha=0.6$, $\kappa_0=2$, $m=0.25$, $\eta=0.08$, $k_{\rm EP}=0.5743977$. Lengths are in $c_b/\Omega$, wavenumbers in $\Omega/c_b$ and velocities in $c_b$.}
\end{figure*}

\emph{Trembling near the exceptional point.---}We now follow a pulse built from a single Floquet branch $\beta$, with spectral width $\sigma_k$ about $k_0$, and write its Floquet mode as $e^{-i\omega_\beta(k)t}\xi_\beta(k,t)$, with complex quasifrequency $\omega_\beta$ and periodic $\xi_\beta(k,t+T)=\xi_\beta(k,t)$ whose $a$ component is $\xi_{\beta,a}$.
Each component carries its own phase, which defines a component connection,
\begin{equation}
\begin{aligned}
\mathcal A_{\beta,a}(k,t)&=-\frac{\Imm\!\left[\xi_{\beta,a}^*\,\partial_k\xi_{\beta,a}\right]}{|\xi_{\beta,a}|^2},\\
\Delta Z_{\beta,a\bar a}(t)&=\Delta\mathcal A_{\beta,a\bar a}(k_0,t)+O(\sigma_k),
\end{aligned}
\label{eq:component_connection}
\end{equation}
where $\Delta\mathcal A_{\beta,a\bar a}=\mathcal A_{\beta,a}-\mathcal A_{\beta,\bar a}$ and $\mathcal A_{\beta,\bar a}$ is the energy-weighted connection of the complement.
Just as the slope of the spectral phase sets the delay of an optical pulse, a component phase $\phi_a$ shifts that component by $-\partial_k\phi_a$, so the source velocity depends on differences between the component phase gradients within one Floquet mode.
The envelope position and the group motion are common to all components and cancel at leading order from this narrow-band difference for $\sigma_k\ll k_{\rm EP}-k_0$ over a fixed window; at component zeros we use the weighted moments directly~\cite{SM}.
In this limit the source weights and connection differences repeat with period $T$, and with $\avgT{\cdot}$ the average over one period we call the zero-mean part of the source velocity the trembling velocity and its centered root-mean-square (rms) over one period the trembling-velocity amplitude,
\begin{equation}
v_{\rm tr}=v_{\rm src}-\avgT{v_{\rm src}},\qquad
A_{\rm tr}=\avgT{v_{\rm tr}^2}^{1/2}.
\label{eq:trembling_amplitude}
\end{equation}
At fixed passband detuning and over a fixed window, the sector centroids of the pulse, weighted by their energy fractions, advance on average at the band slope.
Averaging Eq.~\eqref{eq:centroid_identity} over one period then gives, at leading order in $\sigma_k$,
\begin{equation}
\avgT{v_{\rm flux}}+\avgT{v_{\rm src}}=\partial_k\Ree\omega_\beta
\label{eq:period_average_velocity}
\end{equation}
\cite{SM}.
Because $|v_{\rm flux}|\le c_b$ at every instant, Eq.~\eqref{eq:period_average_velocity} implies that the divergence of the band slope near the EP resides in the mean source velocity.

Near the lower gap edge the two resonant Floquet modes coalesce at $k_{\rm EP}$ [Fig.~\ref{fig:Fig2}(a)]~\cite{dembowski2001experimental,heiss2012physics,miri2019exceptional}.
There the one-period propagation matrix of the full field--matter system has a single eigenvector $\xi_0$ for a doubly degenerate Floquet multiplier and a generalized partner $\xi_1$, which form a Jordan chain.
Removing the Floquet factor, and from the partner also its term linear in time, gives the periodic vectors $\xi_0^{\rm per}(t)$ and $\xi_1^{\rm per}(t)$~\cite{SM}.
With $d\equiv k_{\rm EP}-k>0$ on the passband side, the chain yields the square-root expansion
\begin{equation}
\begin{aligned}
\xi_{\beta,a}(k,t)&=\xi_{0,a}^{\rm per}(t)+\beta\sqrt d\,\chi_a(t)+O(d),\\
\mathcal A_{\beta,a}(k,t)&=\beta\frac{\mathcal C_a(t)}{\sqrt d}+O(1),
\end{aligned}
\label{eq:puiseux_connection}
\end{equation}
for $\beta=\pm$, where $\chi$ follows from $\xi_1^{\rm per}$ and the second line holds where $\xi_{0,a}^{\rm per}(t)\ne0$~\cite{SM}.
The mode thus has a finite limit at the EP while its wavenumber derivative diverges, and the sector mismatch inherits that divergence wherever $\mathcal C_a\ne\mathcal C_{\bar a}$.
Hence the leading source velocity is odd in $\beta$, and when its centered waveform does not vanish identically the trembling-velocity amplitude on either branch obeys
\begin{equation}
A_{\rm tr}=\frac{C_s}{\sqrt d}+O(1),\qquad d\to0^+,
\label{eq:trembling_scaling}
\end{equation}
where $C_s$, the centered rms of the leading source waveform, follows from the propagated chain, the energy weights and the two exchange rates.

The source channels trace a two-humped cycle [Fig.~\ref{fig:Fig2}(b)], and their rms amplitudes follow $d^{-1/2}$ over three decades, with $C_s=0.0822$ for the total source [Fig.~\ref{fig:Fig2}(c)].
At $d=10^{-4}$ the trembling-velocity amplitude reaches $8.1\,c_b$, whereas the flux velocity stays below $0.4\,c_b$ at every plotted detuning.
Of the band slope of $6.3\,c_b$ there, the mean flux velocity carries $0.10\,c_b$ and the mean source velocity $6.2\,c_b$.
In position, the detrended centroid of the packet at $d_0=10^{-4}$, where $d_0=k_{\rm EP}-k_0$ is the detuning of the packet center $k_0$, swings by about $17\,c_b/\Omega$ peak to peak, whereas the flux contribution spans less than $0.06\,c_b/\Omega$ [Fig.~\ref{fig:Fig2}(d)].
Because it integrates the same leading waveform, the displacement amplitude also grows as $d^{-1/2}$~\cite{SM}.
Field measurements can reconstruct the field-energy centroid $Z_{\rm EM}$, weighted by $u_E+u_H$, whose centered rms displacement is about eight times that of $Z$ for the packets of Fig.~\ref{fig:Fig2}(d), and its detrended, centered displacements at the two detunings nearly coincide after multiplication by $\sqrt{d_0}$~\cite{SM}.
In the transmission-line realization, time-resolved line voltages and currents together with resonator charge and current determine all four sector energies and first moments and thereby reconstruct $Z(t)$ and $v_{\rm src}(t)$~\cite{SM}.
The damping contribution, carried by the kinetic-energy sector, exceeds that of modulation in centered rms at $m=0.25$ and $\eta=0.5$~\cite{SM}.

\emph{Nearly parallel modes.---}Near the EP the Jordan structure also makes the two coalescing modes nearly parallel, which the Petermann factor quantifies.
We evaluate it with the field--matter energy metric and its dual,
\begin{equation}
\begin{aligned}
K_\beta^{(u)}
&=\frac{\bra{R_\beta}G_u\ket{R_\beta}\bra{L_\beta}G_u^{-1}\ket{L_\beta}}
{\left|\braket{L_\beta|R_\beta}\right|^2}\\
&=\frac{\mathcal K_0}{d}+O(d^{-1/2}),
\end{aligned}
\label{eq:energy_petermann_factor}
\end{equation}
where $\ket{R_\beta}$ and $\bra{L_\beta}$ are the right and left Floquet eigenvectors, $G_u$ is the period-averaged energy Gramian, $G_u^{-1}$ defines the dual norm of the left covector and $\mathcal K_0>0$~\cite{SM}.
Introduced for the excess noise of gain-guided lasers~\cite{petermann1979calculated}, the factor quantifies mode nonorthogonality~\cite{siegman1989excess,berry2003mode} and diverges at EPs~\cite{wiersig2023petermann,PhysRevResearch.6.013044}, including the momentum-gap edges of PTCs~\cite{park2025spontaneous}.
The same Jordan chain controls the weighted squared sector separations.
We add the inter-sector variances of the two overlapping source-selected partitions, $S_{\rm sep}=f_P f_{\bar P}(\Delta Z_{P\bar P})^2+f_Q f_{\bar Q}(\Delta Z_{Q\bar Q})^2$, and call $A_{\rm sep}$ its centered rms over one period.
Taking the narrow-band limit at fixed $d$ and then approaching the EP gives $A_{\rm sep}=C_{\rm sep}/d+o(d^{-1})$ whenever the leading centered coefficient $C_{\rm sep}$ is nonzero, with a leading term even in $\beta$~\cite{SM} [Fig.~\ref{fig:Fig2}(e)].
Under the stated nonzero-coefficient conditions, the common Jordan chain gives the leading local passband laws
\begin{equation}
\begin{aligned}
A_{\rm tr}&\propto\bigl[K_\beta^{(u)}\bigr]^{1/2}\propto d^{-1/2},\\
A_{\rm sep}&\propto K_\beta^{(u)}\propto d^{-1}.
\end{aligned}
\label{eq:shared_ep_scalings}
\end{equation}

The coefficient $C_s$ contains the energy-exchange rates, whereas $C_{\rm sep}$ is set by the sector fractions and separations.
The Petermann factor alone therefore fixes neither.

\begin{figure}[t]
\centering
\includegraphics[width=\columnwidth]{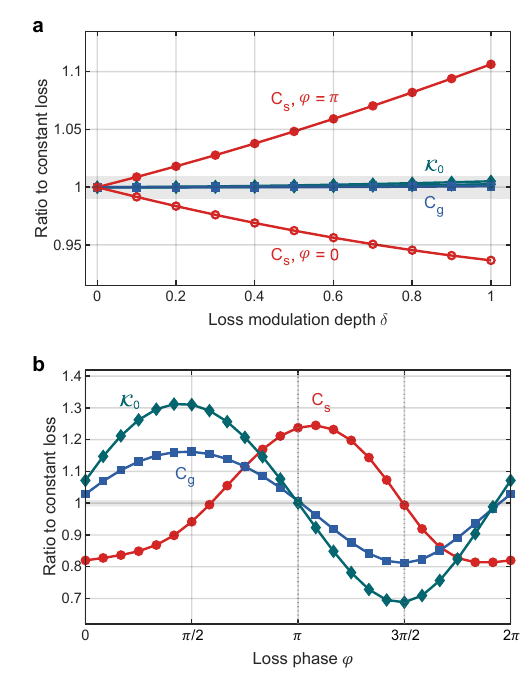}
\caption{\label{fig:Fig_loss}
Loss modulation separates the trembling coefficient from the band-slope and Petermann coefficients.
Leading coefficients $C_s$, $C_g$ and $\mathcal K_0$ at the EP of the loss profile $\eta(t)=\eta_0[1+\delta\cos(\Omega t+\varphi)]$, divided by their values for constant loss $\eta_0$.
(a) Depth dependence for $\eta_0=0.08$ with loss in phase ($\varphi=0$, open symbols) and in antiphase ($\varphi=\pi$, filled symbols) with the restoring coefficient.
(b) Phase dependence for $\eta_0=0.3$ and $\delta=1$; dotted lines, $\varphi=\pi$ and $3\pi/2$.
Shaded bands mark $\pm1\%$ about the constant-loss value. Symbols mark computed points, joined by straight segments. Other parameters as in Fig.~\ref{fig:Fig2}.}
\end{figure}

A common control can change both the restoring coefficient and the damping, and their linear responses are in phase or in antiphase when the relative response lag is negligible.
We take $\eta(t)=\eta_0[1+\delta\cos(\Omega t+\varphi)]$, for which each loss profile has its own EP, and compare $C_s$ with $\mathcal K_0$ and with the band-slope coefficient $C_g=\lim_{d\to0^+}\sqrt d\,|\partial_k\Ree\omega_\beta|$ (Fig.~\ref{fig:Fig_loss}).
For the medium of Fig.~\ref{fig:Fig2}, in-phase ($\varphi=0$) and antiphase ($\varphi=\pi$) loss modulation change $C_s$ while $C_g$ and $\mathcal K_0$ remain close to their constant-loss values [Fig.~\ref{fig:Fig_loss}(a)].
At stronger damping, antiphase loss raises $C_s$ by about a quarter with little change in the other two coefficients, while at $\varphi=3\pi/2$ $C_g$ and $\mathcal K_0$ decrease and $C_s$ remains close to its constant-loss value [Fig.~\ref{fig:Fig_loss}(b)].
At $\eta_0=0.08$, for single-branch packets with $d_0=10^{-4}$, full-depth antiphase loss raises the centered rms displacement of $Z$ by about a tenth and in-phase loss lowers it by about a twentieth, while that of $Z_{\rm EM}$ barely changes~\cite{SM}.
The ratios $C_s/C_g$ and $C_s/\sqrt{\mathcal K_0}$ therefore depend on the loss profile.

\begin{figure}[t]
\centering
\includegraphics[width=\columnwidth]{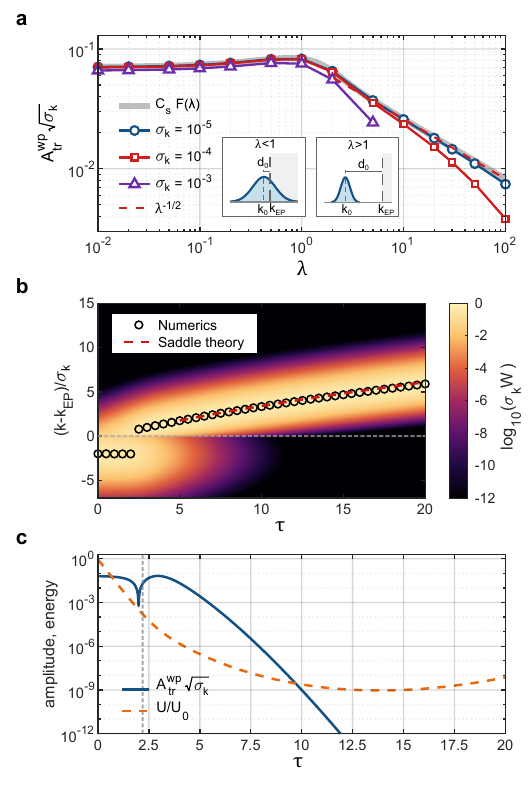}
\caption{\label{fig:Fig3}
Finite bandwidth rounds the trembling, and the spectrum migrates into the gap.
(a) $A_{\rm tr}^{\rm wp}\sqrt{\sigma_k}$ over the first period for single-branch packets with $\sigma_k=10^{-5}$, $10^{-4}$ and $10^{-3}$, against $\lambda=d_0/\sigma_k$.
Gray, Gaussian average $C_s F(\lambda)$; dashed, $\lambda^{-1/2}$.
Insets: spectra overlapping and resolved from the EP.
(b) Normalized spectral energy density $\log_{10}(\sigma_k W)$, with $\int W\,\mathrm{d}k=1$, of the propagated packet with $\sigma_k=10^{-4}$ and $\lambda=2$, against $\tau=t\sqrt{\rho\sigma_k}$.
Circles, spectral maxima; dashed, saddle of Eq.~\eqref{eq:spectral_saddle}; dotted, $k=k_{\rm EP}$.
(c) Scaled trembling amplitude $A_{\rm tr}^{\rm wp}\sqrt{\sigma_k}$, with $A_{\rm tr}^{\rm wp}$ evaluated as the centered rms of $v_{\rm src}$ over each modulation period, and energy $U/U_0$ relative to the launch energy, for the same packet; the dotted line marks the crossing of the spectral maximum into the gap.
Parameters as in Fig.~\ref{fig:Fig2}.}
\end{figure}

\begin{figure*}[t]
\centering
\includegraphics[width=\textwidth]{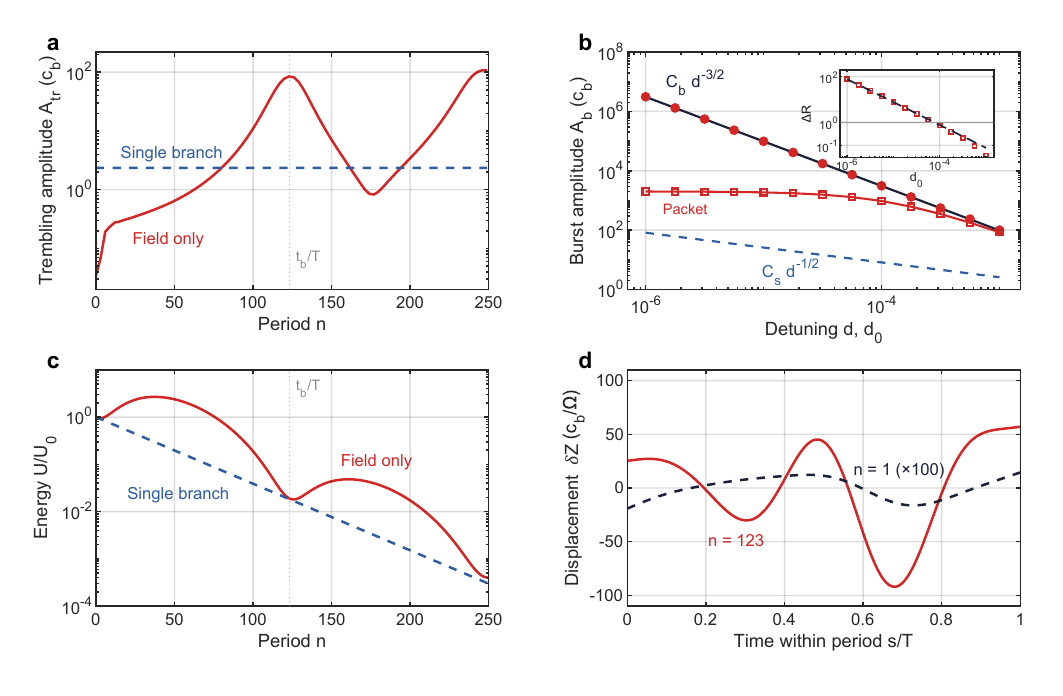}
\caption{\label{fig:Fig_field}
A field-only launch trembles in bursts near the beat recurrence.
(a) Period-resolved $A_{\rm tr}$, the centered rms of $v_{\rm src}$ over one modulation period, sampled every third period, for a forward field pulse with the polarization at rest (field only) and for a single-branch packet with the same envelope, $d_0=10^{-3}$ and $\sigma_k=d_0/20$; dotted line, $t_b/T$.
(b) Burst amplitude $A_b$, the largest one-period centered rms of $v_{\rm src}$ near the first recurrence, for the monochromatic launch (circles) and packets with $\sigma_k=d_0/20$ (squares); solid line, $C_b d^{-3/2}$; dashed line, the single-branch law $C_s d^{-1/2}$.
 Finite packets approach this response when the recurrence-energy excess $\Delta R\simeq0.031\,\sigma_k^2/d_0^3$ is small. Inset, $\Delta R$, the excess over one of the launch-normalized energy ratio of the two packets at the recurrence minimum (squares), and the estimate (dashed); gray line, $\Delta R=1$.
(c) Energy $U/U_0$ relative to the launch energy for the two packets of (a).
(d) Centroid displacement $\delta Z$ of the field-only packet, with a least-squares linear fit over each period removed, within the burst period $n=123$ and within $n=1$, the latter multiplied by $100$.
Parameters as in Fig.~\ref{fig:Fig2}.}
\end{figure*}

\emph{Finite bandwidth, gap migration and fade.---}We consider a Gaussian pulse of rms spectral width $\sigma_k$ centered at $k_0$ on the passband side, with $\lambda=d_0/\sigma_k$.
For a regular launch with a real envelope in our eigenvector gauge and over a fixed window, the leading passband source waveform is common to all wavenumbers near the EP, and averaging it over the spectrum gives, at fixed $\lambda$,
\begin{equation}
\begin{aligned}
A_{\rm tr}^{\rm wp}\sqrt{\sigma_k}&=C_s F(\lambda)+o(1),\qquad\sigma_k\to0,\\
F(\lambda)&=\int_0^\infty\frac{e^{-(u-\lambda)^2/2}}{\sqrt{2\pi u}}\,\mathrm{d}u,
\end{aligned}
\label{eq:finite_bandwidth_scaling}
\end{equation}
with $F(0)=0.860$ and $F\to\lambda^{-1/2}$ for $\lambda\gg1$~\cite{SM}.
A finite bandwidth thus rounds the local divergence into a response of order $C_s\sigma_k^{-1/2}$ when the pulse overlaps the EP.
We propagate the pulses without a narrow-band expansion and launch each eigenvector with its electric component real and positive, a choice that fixes the phase across the EP~\cite{SM}.
For $\sigma_k=10^{-5}$ and $10^{-4}$ the scaled amplitudes $A_{\rm tr}^{\rm wp}\sqrt{\sigma_k}$ approach $C_s F(\lambda)$ and depart from it progressively as the pulse samples wavenumbers beyond the square-root regime [Fig.~\ref{fig:Fig3}(a)], whereas for the broader pulse with $\sigma_k=10^{-3}$ the response already departs from this form at $d_0=0$~\cite{SM}.

On the gap side, $h=k-k_{\rm EP}>0$, the larger Floquet growth rate $\Gamma(k)=\Imm\omega(k)$ rises as $\Gamma(k)-\Gamma_{\rm EP}\simeq\sqrt{\rho h}$, with $\rho=0.0158$ for this medium, and with $y=h/\sigma_k$ this relative gain competes with the Gaussian penalty away from $k_0$, so the spectral weight migrates.
Once the gap-side contribution dominates, its spectral maximum follows the saddle
\begin{equation}
\sqrt{y_s}\,(y_s+\lambda)=\tau,\qquad\tau=t\sqrt{\rho\sigma_k},
\label{eq:spectral_saddle}
\end{equation}
while the local growth expansion holds, with $y_s\simeq\tau^{2/3}$ once $y_s\gg|\lambda|$.

For $\lambda=2$ the maximum of the propagated spectrum stays on the passband until $\tau\approx2.2$, jumps into the gap and then follows Eq.~\eqref{eq:spectral_saddle} [Fig.~\ref{fig:Fig3}(b)].
Because $\Gamma_{\rm EP}<0$, the packet first loses energy while its normalized spectrum shifts into the gap, passes through an energy minimum near the stage when the saddle reaches the leading net-growth scale $h\simeq\Gamma_{\rm EP}^2/\rho\approx4\times10^{-4}$ [Fig.~\ref{fig:Fig3}(c)], and then recovers~\cite{SM}.

The trembling fades as the gap side takes over [Fig.~\ref{fig:Fig3}(c)] because of a symmetry of the gap modes.
In the frame $(E,-iH,P,Q)$ the equations of motion are real.
On a smooth gap branch with a nondegenerate negative real Floquet multiplier, our launch convention therefore makes the propagated state real in that frame, so the gap-side contributions to every first-moment integral vanish~\cite{SM}.
Hence the trembling decays as the passband contributions to the normalized moments shrink.
A spectral phase varying nonlinearly with $k$ can shift the sector centroids unequally and lift this gap-side null.
A quadratic chirp keeps a finite gap-side trembling over the computed window.
The real-envelope response instead falls by nearly eight orders of magnitude by $\tau=10$, and the two smoothed finite-variance launches follow this decrease through $\tau=10$~\cite{SM}.

\emph{Launch in the fields alone.---}An eigen-packet occupies one Floquet branch, whereas a forward field pulse launched with the polarization at rest, $x(k,0)=\psi(k)(1,1,0,0)^{\rm T}$ for $\epsilon_b=\mu=1$, projects onto both coalescing modes.
Because the modes are nearly parallel, their branch-diagonal energies greatly exceed the launch energy and are balanced by destructive interference, and their singular source contributions cancel in the same way, so the first-period trembling remains finite as the EP is approached.
The two branches split as $\Ree\omega_\pm\simeq\Ree\omega_{\rm EP}\pm2C_g\sqrt d$, giving the beat time $t_b\simeq\pi/(2C_g\sqrt d)$.
Near this recurrence the energy reaches an interference minimum while the two branch components have separated by about $\pi/d$, their first-moment contributions no longer cancel, and the trembling bursts [Fig.~\ref{fig:Fig_field}(a)].
In the monochromatic limit of this preparation the computed burst amplitudes follow $C_bd^{-3/2}$ with $C_b\simeq0.0031$ over $10^{-6}\le d\le10^{-3}$ and exceed the single-branch amplitude [Fig.~\ref{fig:Fig_field}(b)].
A finite spectral width spreads the relative beat phase across the pulse, the large branch-diagonal energies amplify its effect, and the burst approaches its monochromatic value only while the resulting energy excess at the recurrence minimum, $\Delta R\simeq0.031\,\sigma_k^2/d_0^3$, stays small, a condition that fails at fixed $\sigma_k/d_0$ as the EP is approached [Fig.~\ref{fig:Fig_field}(b), inset]~\cite{SM}.
For $d_0=10^{-3}$ and $\sigma_k=d_0/20$ the burst peaks about $123$ periods after launch, when the pulse retains about $2\%$ of its launch energy [Fig.~\ref{fig:Fig_field}(c)], and the detrended centroid swings by about $150\,c_b/\Omega$ peak to peak, a small fraction of the initial rms width $1/(2\sigma_k)$ [Fig.~\ref{fig:Fig_field}(d)].

\emph{Discussion.---}Near a PTC gap edge, growing offsets between energy sectors amplify the source contribution to the energy-centroid velocity while the Poynting-flux velocity stays bounded.
In the local single-branch passband limit the source contribution has a period average that carries the divergent part of the band slope and a zero-mean part that produces intracycle trembling.
Cycle averaging therefore removes this trembling velocity.
Neither the band slope nor the Petermann factor fixes its amplitude: we find that loss modulated in phase or in antiphase with the restoring coefficient can change the trembling coefficient while both leading coefficients stay essentially fixed, and that the preparation decides whether the enhanced response appears at once, fades as gap modes dominate, or recurs in bursts near the beat time.
Beyond the Maxwell--Lorentz medium, the centroid identity applies to modulated linear wave systems whose stored energy has a positive, wavenumber-independent metric in a fixed physical basis~\cite{SM}.

\FloatBarrier
\begin{acknowledgements}
\emph{Acknowledgements.—}
This work is supported by the National Research Foundation of Korea (NRF) through the government of Korea (RS-2022-NR070636) and the Samsung Science and Technology Foundation (SSTF-BA240202). K.L. also acknowledges support from the KAIST Jang Young Sil Fellow Program.
\end{acknowledgements}

\bibliography{ptc_ZB}

\clearpage
\onecolumngrid
\setcounter{page}{1}
\renewcommand{\thepage}{S\arabic{page}}
\setcounter{equation}{0}
\renewcommand{\theequation}{S\arabic{equation}}
\providecommand{\theHequation}{}
\renewcommand{\theHequation}{supp.\arabic{equation}}
\setcounter{section}{0}
\setcounter{secnumdepth}{1}
\renewcommand{\thesection}{S\arabic{section}}
\providecommand{\theHsection}{}
\renewcommand{\theHsection}{supp.\arabic{section}}
\setcounter{figure}{0}
\renewcommand{\thefigure}{S\arabic{figure}}
\providecommand{\theHfigure}{}
\renewcommand{\theHfigure}{supp.\arabic{figure}}

\begin{center}
{\large\bfseries Supplemental Material: Energy-Centroid Trembling near Exceptional Points in Photonic Time Crystals\par}
\medskip
Kyungmin Lee$^1$, Younsung Kim$^1$, Minwook Kyung$^1$, Seungmin Yeon$^1$, Bumki Min$^{1,\dagger}$\par
\smallskip
{\small\itshape $^1$Department of Physics, Korea Advanced Institute of Science and Technology, Daejeon 34141, Republic of Korea\par}
\smallskip
{\small $^\dagger$\href{mailto:bmin@kaist.ac.kr}{bmin@kaist.ac.kr}\par}
\end{center}

\section{Model and conventions}
\label{sec:sm_model}
We take $\mathbf E=E\hat{\mathbf x}$, $\mathbf H=H\hat{\mathbf y}$, $\mathbf P=P\hat{\mathbf x}$, $D=\epsilon_b E+P$ and $B=\mu H$ with $\epsilon_b,\mu>0$, and the driven Lorentz oscillator
\begin{equation}
\ddot P+\eta\dot P+\kappa(t)P=\alpha E,\qquad
\alpha>0,\quad\eta\ge0,\quad
\kappa(t)=\kappa_0[1+m\cos\Omega t]>0,\quad Q\equiv\dot P.
\label{eq:sm_lorentz_conventions}
\end{equation}
The model describes oscillators of fixed density with a prescribed stiffness $\kappa(t)$ and a local viscous loss $\eta(t)$.
For bound charges $q$ of mass $M$ and density $N$ in the potential $k(t)x^2/2$,
\begin{equation}
P=Nqx,\qquad
\alpha=\frac{Nq^2}{M},\qquad
\kappa(t)=\frac{k(t)}{M},
\label{eq:sm_microscopic}
\end{equation}
and $u_P+u_Q$ represents the instantaneous mechanical energy of the oscillators in the complex-field normalization used here.
The centroid $Z$ therefore locates the energy held by the fields and oscillators, while the pump work and the dissipated heat enter through the source terms of Eq.~\eqref{eq:energy_balance}.

In the continuum limit the same equations describe a transmission line with inductance $L'$ and capacitance $C'$ per length, shunt-loaded at density $\rho_r$ by series resonators with inductance $L_r$, resistance $R(t)$ and capacitance $C_r(t)$.
With the line voltage and current in place of $E$ and $H$ and $P=\rho_rq_r$ for the resonator charge $q_r$,
\begin{equation}
\epsilon_b=C',\qquad
\mu=L',\qquad
\alpha=\frac{\rho_r}{L_r},\qquad
\kappa(t)=\frac{1}{L_rC_r(t)},\qquad
\eta(t)=\frac{R(t)}{L_r},
\label{eq:sm_circuit_map}
\end{equation}
and $u_{\rm tot}$ becomes the line and resonator energy.
A conductance in parallel with $C_r(t)$ gives a different equation, so $\eta(t)$ describes series loss.
For a varactor, this correspondence applies to linear signals about a prescribed pump trajectory when bias-network loading, pump feedthrough, device memory, parasitic and lattice corrections, and signal-dependent series-resistance terms proportional to the pump current are negligible.
Cell-resolved measurements give the sector energies directly.
For a cell $j$ of length $\Delta z$ at position $z_j$,
\begin{equation}
U_{E,j}=\frac{C'\Delta z}{2}V_j^2,\qquad
U_{H,j}=\frac{L'\Delta z}{2}I_j^2,\qquad
U_{P,j}=\frac{q_j^2}{2C_r(t)},\qquad
U_{Q,j}=\frac{L_r}{2}i_{r,j}^2,
\label{eq:sm_circuit_energies}
\end{equation}
with $q_j=C_r(t)v_{C,j}$ from the capacitor voltage and $i_{r,j}=\dot q_j$ the resonator current, summed over the resonators of a cell.
The sums $U_a=\sum_jU_{a,j}$ and $J_a=\sum_jz_jU_{a,j}$ give $Z$ and the sector centroids, and Eq.~\eqref{eq:sm_linear_source} gives $v_{\rm src}$.
The reconstruction uses signals linearized about the pump trajectory, calibrated incremental circuit parameters, and a measured section that contains the pulse.
With complex signals the squares become squared moduli and the factors $1/2$ become $1/4$.
For the numerical results we set $\epsilon_b=\mu=\Omega=1$; time, frequency, wavenumber (including $d$ and $\sigma_k$), length and velocity are expressed in units of $1/\Omega$, $\Omega$, $\Omega/c_b$, $c_b/\Omega$ and $c_b$, respectively, and the quoted coefficients $C_s$, $C_b$, $C_g$ and $\mathcal K_0$ are dimensionless in these units.

Fields are written $X(z,t)=(2\pi)^{-1}\int X(k,t)e^{ikz}\,\mathrm{d}k$, so $z\leftrightarrow i\partial_k$, and time harmonics as $e^{-i\omega t}$.
With complex amplitudes, $u_{\rm EM}=(\epsilon_b|E|^2+\mu|H|^2)/4$ and $S_z=\Ree(EH^*)/2$ are the averages of the values for the real and imaginary parts.

For a pulse whose spectrum lies at $k>0$, these averages also give the energies and first moments of the single real field $\Ree X$ at every instant.
With $\Ree X=(X+X^*)/2$ the cross terms obey
\begin{equation}
\int X^2\,\mathrm{d}z=\frac{1}{2\pi}\int X(k)X(-k)\,\mathrm{d}k=0,\qquad
\int zX^2\,\mathrm{d}z=\frac{i}{2\pi}\int X(-k)\,\partial_kX(k)\,\mathrm{d}k=0,
\label{eq:sm_real_field}
\end{equation}
since the two factors in each integrand have disjoint supports when the spectrum avoids $k\le0$.
The sector energies, first moments and $Z$ of the real pulse therefore equal those of the complex description, while the local densities differ by terms that oscillate at about $2k_0$ in $z$.
For the untruncated Gaussian launches used here this correspondence holds up to negative-wavenumber tails of relative weight of order $\exp(-k_0^2/2\sigma_k^2)$.
Reconstructing $Z(t)$ from measurements needs a time resolution finer than the modulation period.

At fixed $k$ the state $x=(E,H,P,Q)^{\mathrm T}$ obeys
\begin{equation}
\dot x=\mathsf A(k,t)x,\qquad
\mathsf A=\begin{pmatrix}
0&-ik/\epsilon_b&0&-1/\epsilon_b\\
-ik/\mu&0&0&0\\
0&0&0&1\\
\alpha&0&-\kappa(t)&-\eta
\end{pmatrix}.
\label{eq:sm_state_matrix}
\end{equation}

For $\kappa=\kappa_0$ the medium has $\epsilon(\omega)=\epsilon_b+\alpha/(\kappa_0-\omega^2-i\eta\omega)$, and $k^2=\mu\epsilon(\omega)\omega^2$ gives the static Lorentz branches.
We call the one-period propagator $M(k)=U_k(T,0)$ the monodromy and write its eigenvalues, the Floquet multipliers, as $\zeta_\beta=e^{-i\omega_\beta T}$.
A Floquet mode is $x=e^{-i\omega_\beta t}\xi_\beta(k,t)$ with $T$-periodic $\xi_\beta$.
Harmonic-space (Sambe) right and left eigenvectors are written $\ket R$, $\bra L$.
The Floquet calculations underlying Fig.~\ref{fig:Fig2} use a Sambe truncation with harmonics $n=-6,\ldots,6$, the Petermann factor for Fig.~\ref{fig:Fig_loss} uses the time-domain representation of Sec.~\ref{sec:sm_separations_petermann}, and the packet modes of Fig.~\ref{fig:Fig3} are obtained from the time-domain monodromy. Packet dynamics are propagated in the time domain as described in Sec.~\ref{sec:sm_numerical_propagation}. The two-band illustration of Sec.~\ref{sec:sm_floquet_ep} uses static eigenvectors.

\section{Energy balance and centroid identity}
\label{sec:sm_energy_centroid}
With $u_{\rm mat}=(|Q|^2+\kappa|P|^2)/(4\alpha)$ the sector balances are
\begin{equation}
\partial_t u_{\rm EM}+\partial_z S_z=-\frac12\Ree(EQ^*),\qquad
\partial_t u_{\rm mat}=\frac12\Ree(EQ^*)-\frac{\eta}{2\alpha}|Q|^2
+\frac{\dot\kappa}{4\alpha}|P|^2.
\label{eq:sm_sector_balances}
\end{equation}
Their sum is the exact local balance
\begin{equation}
\partial_t u_{\rm tot}+\partial_z S_z=s=\frac{\dot\kappa}{\kappa}u_P-2\eta u_Q,\qquad
u_P=\frac{\kappa|P|^2}{4\alpha},\quad u_Q=\frac{|Q|^2}{4\alpha},
\label{eq:sm_local_balance}
\end{equation}
with $u_{\rm tot}=u_E+u_H+u_P+u_Q\ge0$.
Pointwise $|S_z|\le|E||H|/2\le c_b u_{\rm EM}\le c_b u_{\rm tot}$, so $|v_{\rm flux}|\le c_b$.

For a localized pulse with $S_z$ and $zS_z$ vanishing at the boundaries, multiplying Eq.~\eqref{eq:sm_local_balance} by $z$ and integrating give
\begin{equation}
\dot Z=\frac1U\int S_z\,\mathrm{d}z+\frac1U\int(z-Z)s\,\mathrm{d}z\equiv v_{\rm flux}+v_{\rm src}.
\label{eq:sm_centroid_balance}
\end{equation}
With $U_a=\int u_a\,\mathrm{d}z$, $f_a=U_a/U$, $Z_a=U_a^{-1}\int z u_a\,\mathrm{d}z$ and the complement $\bar a$, $Z=f_a Z_a+f_{\bar a}Z_{\bar a}$, so $U^{-1}\int(z-Z)u_a\,\mathrm{d}z=f_a f_{\bar a}(Z_a-Z_{\bar a})$, which gives Eq.~\eqref{eq:centroid_identity}.
For any partition $a|\bar a$ the energy variance splits as
\begin{equation}
\sigma_z^2=f_a\sigma_a^2+f_{\bar a}\sigma_{\bar a}^2
+f_a f_{\bar a}(\Delta Z_{a\bar a})^2,\qquad
|f_a f_{\bar a}\Delta Z_{a\bar a}|\le\sqrt{f_a f_{\bar a}}\,\sigma_z\le\sigma_z/2,
\label{eq:sm_partition_variance}
\end{equation}
where $\sigma_a$ and $\sigma_{\bar a}$ are the internal widths of the two parts, which gives Eq.~\eqref{eq:source_width_bound}.

The identity extends beyond the Maxwell--Lorentz medium to any evolution $\partial_t x=-i\mathsf H(k,t)x$ in a fixed physical basis with stored energy $U=x^\dagger Gx$, where $G(t)$ is positive and independent of $k$, and with centroid $Z=\Ree(ix^\dagger G\partial_k x)/U$.
Then $\dot U=x^\dagger\Pi x$ with the power operator $\Pi=i(\mathsf H^\dagger G-G\mathsf H)+\dot G$, and
\begin{equation}
\dot Z=\frac{\Ree\!\left[x^\dagger G(\partial_k \mathsf H)x\right]}{U}
+\frac{\Ree(ix^\dagger\Pi\partial_k x)-Z\,x^\dagger\Pi x}{U}.
\label{eq:sm_positive_metric_centroid}
\end{equation}
A common gain $\Pi\to\Pi+c(t)G$ with real $c(t)$ leaves the second term unchanged.
For a finite packet the numerator and denominator are integrated over $k$, and the moments must exist.
If $G=\sum_a G_a$ with nonnegative, $k$-independent sector matrices $G_a$ that define the sector energies and moments, and $\Pi=\sum_a r_a(t)G_a$, the source term splits into the corresponding sector channels.
For the Maxwell--Lorentz medium, $G=\operatorname{diag}(\epsilon_b,\mu,\kappa/\alpha,1/\alpha)/4$ and $\Pi=\operatorname{diag}(0,0,\dot\kappa/(4\alpha),-2\eta/(4\alpha))$, which recovers Eq.~\eqref{eq:centroid_identity}.

\section{Packets, connections and the gap-side null}
\label{sec:sm_packets_gap_null}
To express the sector centroids through the spectral amplitudes, write each sector density as $u_a=c_a(t)|X_a|^2$ with $(c_E,c_H,c_P,c_Q)=(\epsilon_b,\mu,\kappa/\alpha,1/\alpha)/4$.
With $X_a(k,t)=\mathcal F_a(k,t)$, Parseval's identity gives
\begin{equation}
Z_a=\frac{\Ree\int \mathrm{d}k\,\mathcal F_a^*\,i\partial_k\mathcal F_a}
{\int \mathrm{d}k\,|\mathcal F_a|^2}.
\label{eq:sm_parseval_centroid}
\end{equation}
Its denominator times $c_a/(2\pi)$ is the sector energy $U_a$, and $J_a=U_aZ_a$ is the sector first moment. The identity $z\leftrightarrow i\partial_k$ thus expresses each sector centroid as an average of $-\partial_k\arg\mathcal F_a$, weighted by that component's spectral energy.
For a single branch, $\mathcal F_{\beta,a}=\psi(k)\xi_{\beta,a}(k,t)e^{-i\omega_\beta t}$ with $\omega_\beta=\omega_{\beta,R}+i\Gamma_\beta$.
With $\psi=|\psi|e^{i\phi_\psi}$ we define the envelope connection, the component connection and the component weight,
\begin{equation}
\mathcal A_{\rm env}=-\partial_k\phi_\psi,\qquad
\mathcal A_{\beta,a}=-\frac{\Imm[\xi_{\beta,a}^*\partial_k\xi_{\beta,a}]}{|\xi_{\beta,a}|^2},\qquad
p_{\beta,a}=|\psi|^2|\xi_{\beta,a}|^2e^{2\Gamma_\beta t},
\label{eq:sm_connections_weights}
\end{equation}
and Eq.~\eqref{eq:sm_parseval_centroid} then gives
\begin{equation}
Z_{\beta,a}^{\rm wp}(t)=\left\langle\mathcal A_{\rm env}+t\,\partial_k\omega_{\beta,R}+\mathcal A_{\beta,a}\right\rangle_{\beta,a,t},
\label{eq:sm_exact_sector_centroid}
\end{equation}
where $\langle\cdot\rangle_{\beta,a,t}$ averages with $p_{\beta,a}$.
Differences $Z_{\beta,a}^{\rm wp}-Z_{\beta,j}^{\rm wp}$ contain the common terms averaged with different weights, so they do not reduce to differences of local connections at finite bandwidth.
In this notation the fractions and sector separations are
\begin{equation}
f_{\beta,a}^{\rm wp}=\frac{c_a\int p_{\beta,a}\,\mathrm{d}k}{\sum_j c_j\int p_{\beta,j}\,\mathrm{d}k},\qquad
\Delta Z_{\beta,a\bar a}^{\rm wp}=\frac{1}{f_{\beta,\bar a}^{\rm wp}}\sum_{j\ne a}f_{\beta,j}^{\rm wp}\bigl(Z_{\beta,a}^{\rm wp}-Z_{\beta,j}^{\rm wp}\bigr).
\label{eq:sm_packet_fractions_separations}
\end{equation}

Under the gauge change $\xi\to e^{i\gamma(k)}\xi$ and $\psi\to e^{-i\gamma(k)}\psi$ the amplitude $\mathcal F$ is unchanged, $\mathcal A_{\rm env}\to\mathcal A_{\rm env}+\partial_k\gamma$ and $\mathcal A_{\beta,a}\to\mathcal A_{\beta,a}-\partial_k\gamma$, so the sum $\mathcal A_{\rm env}+\mathcal A_{\beta,a}$ and all connection differences are gauge invariant.

The narrow-band limit recovers the main-text relations, because at fixed regular $k_0$ on the passband, over a fixed window and for $\sigma_k\ll|k_{\rm EP}-k_0|$, $\langle F\rangle_{\beta,a,t}=F(k_0,t)+O(\sigma_k)$.
In the sector differences the common terms cancel at leading order, leaving $\Delta\mathcal A_{\beta,a\bar a}(k_0,t)$ as the leading sector-to-complement separation, with the complement connection $\mathcal A_{\beta,\bar a}=\sum_{j\ne a}f_{\beta,j}\mathcal A_{\beta,j}/f_{\beta,\bar a}$ and $f_{\beta,a}=c_a|\xi_{\beta,a}|^2/\sum_j c_j|\xi_{\beta,j}|^2$ at $k_0$.
Summing Eq.~\eqref{eq:sm_exact_sector_centroid} with the fractions gives
\begin{equation}
Z(t)=\mathcal A_{\rm env}(k_0)+t\,\partial_k\omega_{\beta,R}(k_0)
+\sum_a f_{\beta,a}\mathcal A_{\beta,a}(k_0,t)+O(\sigma_k).
\label{eq:sm_narrowband_total_centroid}
\end{equation}
Its connection part $\sum_a f_{\beta,a}\mathcal A_{\beta,a}$ is $T$-periodic, whereas the finite-bandwidth remainder need not be.
For the Gaussian preparation, differential decay across the spectrum shifts the spectral center by $O(\sigma_k^2t)$, and averaging Eq.~\eqref{eq:sm_centroid_balance} over one period then gives Eq.~\eqref{eq:period_average_velocity} at leading order.
The error constants depend on $k_0$ and need not stay bounded as $k_0\to k_{\rm EP}$, where the joint limit of Sec.~\ref{sec:sm_finite_bandwidth} applies.

At component nodes the connections of Eq.~\eqref{eq:sm_connections_weights} are undefined, so to avoid dividing by component intensities, define
\begin{equation}
\begin{aligned}
&n_{\beta,a}=c_a|\xi_{\beta,a}|^2,\qquad
N_\beta=\sum_a n_{\beta,a}>0,\qquad
\mathcal J_{\beta,a}=-c_a\Imm(\xi_{\beta,a}^*\partial_k\xi_{\beta,a}),\\
&\mathcal L_{\beta,a}=\frac{\mathcal J_{\beta,a}-f_{\beta,a}\sum_j\mathcal J_{\beta,j}}{N_\beta}
=f_{\beta,a}f_{\beta,\bar a}\Delta\mathcal A_{\beta,a\bar a}.
\end{aligned}
\label{eq:sm_node_safe_moments}
\end{equation}
The first expression for $\mathcal L_{\beta,a}$ stays defined at component nodes, and the local source velocity is
\begin{equation}
v_{{\rm src},\beta}^{\rm loc}=\frac{\dot\kappa}{\kappa}\mathcal L_{\beta,P}-2\eta\mathcal L_{\beta,Q}.
\label{eq:sm_local_source}
\end{equation}
For a state that mixes branches, as in Sec.~\ref{sec:sm_field_launch}, the moments of Eq.~\eqref{eq:sm_node_safe_moments} are evaluated from the full propagated state and its wavenumber derivative at fixed time.
The amplitudes $A_{\rm tr}$ and $A_b$ are centered rms values of $v_{\rm src}$, and the total centroid velocity adds $v_{\rm flux}$, with $|v_{\rm flux}|\le c_b$.

The gap-side null follows from an antilinear symmetry at real $k$: with $\mathcal P=\operatorname{diag}(1,-1,1,1)$,
\begin{equation}
\mathcal P\mathsf A^*\mathcal P=\mathsf A,\qquad
\mathcal P U_k^*\mathcal P=U_k,\qquad
\mathcal P M^*\mathcal P=M,
\label{eq:sm_antilinear_symmetry}
\end{equation}
so the equations of motion are real in the frame $(E,-iH,P,Q)$.
Take a smooth branch with a nondegenerate negative real multiplier $\zeta_\beta$ on an open $k$-interval inside the gap, with $\omega_\beta=\Omega/2+i\Gamma_\beta$.
Simplicity gives $\mathcal P\ket{r_\beta}^*=e^{i\theta_\beta(k)}\ket{r_\beta}$ with unit-modulus factor, and propagation gives
\begin{equation}
\xi_{\beta,a}=e^{i(\Omega t-\theta_\beta)/2}s_a x_{\beta,a},\qquad
(s_E,s_H,s_P,s_Q)=(1,i,1,1),\qquad x_{\beta,a}\in\mathbb R.
\label{eq:sm_gap_mode_real_frame}
\end{equation}
Then $\mathcal J_{\beta,a}=(\partial_k\theta_\beta)n_{\beta,a}/2$ and $\mathcal L_{\beta,a}=0$, also at nodes, so the local source velocity and the local squared separation vanish channel by channel for either sign of $\Gamma_\beta$.
For a packet on this branch, Eq.~\eqref{eq:sm_exact_sector_centroid} gives
\begin{equation}
Z_{\beta,a}^{\rm wp}=\langle\mathcal G_\beta\rangle_{\beta,a,t},\qquad
\mathcal G_\beta(k)=\mathcal A_{\rm env}(k)+\tfrac12\partial_k\theta_\beta(k),
\label{eq:sm_gap_packet_phase}
\end{equation}
with gauge-invariant $\mathcal G_\beta$.
Unequal sector weights can give unequal averages of a nonconstant $\mathcal G_\beta$.

For the unchirped eigen-packets of the main text we launch each eigenvector with its electric component real and positive wherever $E(k,0)\ne0$, under a real envelope.
On the gap branch this sets $\mathcal G_\beta\equiv0$, so the propagated state is real in the frame $(E,-iH,P,Q)$ up to a $k$-independent factor, the gap-side integrands $\Ree(i\mathcal F_a^*\partial_k\mathcal F_a)$ of Eq.~\eqref{eq:sm_parseval_centroid} vanish, and the gap side adds energy weight to every sector without adding first moment.
Hence the source velocity of the packet comes only from passband contributions to the normalized moments, and their decrease accounts for the fade of Fig.~\ref{fig:Fig3}(c).
We describe the gap side through its contributions to the moment integrals of the full continuous packet.
A packet cut sharply at $k_{\rm EP}$ with nonzero endpoint amplitude has $1/z$ tails and divergent first moments.

To discuss finite-second-moment approximations, consider the launch $\psi_0(k)x(k)$ with the real Gaussian envelope $\psi_0$ and a mode profile $x$ whose spectral completion is regular away from the EP.
Near the EP the profile has $\partial_k x\propto|k-k_{\rm EP}|^{-1/2}$ (Sec.~\ref{sec:sm_floquet_ep}).
First-moment integrands are then integrable, while $\int|\partial_k x|^2\,\mathrm{d}k$ diverges logarithmically, so the ideal eigen-packet has finite centroids and an infinite spatial variance.
Replace the profile by $x_\varepsilon$, equal to $x$ outside an interval of width $\varepsilon_k$ about $k_{\rm EP}$, with no added rapid phase and with $\|x_\varepsilon-x\|=O(\varepsilon_k^{1/2})$ and $\|\partial_k x_\varepsilon\|=O(\varepsilon_k^{-1/2})$ inside it.
With continuous matching at the smoothing boundaries and a square-integrable derivative of the full prepared spectrum outside the smoothing interval, the regularized launch has finite variance. The local smoothing changes its unnormalized first moments by $O(\varepsilon_k^{1/2})$ and its energy by $O(\varepsilon_k^{3/2})$.
At fixed $\lambda$, over a fixed window and for small $\sigma_k$, the packet energy obeys $c\,\sigma_k\le U(t)\le C\,\sigma_k$, so the source velocity changes by $O(\varepsilon_k^{1/2}/\sigma_k)$ and $\sqrt{\sigma_k}A_{\rm tr}^{\rm wp}$ by $O[(\varepsilon_k/\sigma_k)^{1/2}]$.
Therefore the scaled response of Eq.~\eqref{eq:finite_bandwidth_scaling} is unchanged as $\varepsilon_k/\sigma_k\to0$. Section~\ref{sec:sm_finite_variance} implements this regularization numerically.

\section{Floquet modes near the exceptional point}
\label{sec:sm_floquet_ep}
Expanding $\xi_a=\sum_n\xi_{a,n}e^{-in\Omega t}$ gives the connection in harmonic form,
\begin{equation}
\begin{aligned}
\mathcal A_a&=\frac{\mathcal J_a^{\rm diag}+\mathcal J_a^{\rm coh}(t)}
{\mathcal N_a^{\rm diag}+\mathcal N_a^{\rm coh}(t)},\qquad
\mathcal N_a^{\rm coh}=\sum_{n\ne n'}\xi_{a,n}^*\xi_{a,n'}e^{i(n-n')\Omega t},\\
\mathcal J_a^{\rm coh}&=-\Imm\sum_{n\ne n'}\xi_{a,n}^*\partial_k\xi_{a,n'}e^{i(n-n')\Omega t},
\end{aligned}
\label{eq:sm_harmonic_connection}
\end{equation}
with $\mathcal N_a^{\rm diag}=\sum_n|\xi_{a,n}|^2$ and $\mathcal J_a^{\rm diag}=-\Imm\sum_n\xi_{a,n}^*\partial_k\xi_{a,n}$; the common sector weight $c_a$ of Eq.~\eqref{eq:sm_node_safe_moments} cancels between numerator and denominator.
All time dependence of the connection comes from coherence between unequal harmonics.

Near the EP the modes follow from a Jordan chain, since at $k_{\rm EP}$ the monodromy $M_0$ has a defective double multiplier $\zeta_0$ with
\begin{equation}
\begin{aligned}
(M_0-\zeta_0)\xi_0&=0,\qquad
(M_0-\zeta_0)\xi_1=\xi_0,\qquad
\ell^T(M_0-\zeta_0)=0,\\
\ell^T\xi_0&=0,\qquad \ell^T\xi_1=1,
\end{aligned}
\label{eq:sm_jordan_chain}
\end{equation}
and $\xi_1\perp\xi_0$.
Because $M_0$ is real in the frame $(E,-iH,P,Q)$ and $\zeta_0$ is real and negative, $\xi_0$ and $\xi_1$ are real there.
The propagated vectors $a(t)=U_{k_{\rm EP}}(t)\xi_0$ and $b(t)=U_{k_{\rm EP}}(t)\xi_1$ obey $a(t+T)=\zeta_0a(t)$ and $b(t+T)=\zeta_0b(t)+a(t)$, and the periodic chain is
\begin{equation}
\xi_0^{\rm per}(t)=e^{-\Lambda_0t}a(t),\qquad
\xi_1^{\rm per}(t)=e^{-\Lambda_0t}\left[b(t)-\frac{t}{T\zeta_0}a(t)\right],\qquad
\Lambda_0=\frac{\ln|\zeta_0|-i\pi}{T}.
\label{eq:sm_periodic_chain}
\end{equation}
An ordinary Floquet eigenmode $\xi_0^{\rm per}$ exists at the EP, and the generalized partner carries the factor linear in $t$.

For small nonzero detuning $d=k_{\rm EP}-k$ from the EP, with $M(k)=M_0-dM_1$ to first order, the chain yields the square-root expansion: to leading order the half-splitting of the multipliers obeys $\delta\zeta^2=K_Md$ with $K_M=-\ell^TM_1\xi_0$, and the eigenvectors are $v_\pm=\xi_0\pm\delta\zeta\,\xi_1$ with $\zeta_\pm=\zeta_0\pm\delta\zeta$.
We fix the branch of $\delta\zeta$ so that $v_+$ has the larger real quasifrequency.
For the medium of the main text, $\zeta_0=-0.98379$ and $K_M=-0.60339$, so $\delta\zeta=i\sqrt{|K_M|d}$ on the passband side.
On the gap side the same expansion gives the growth-rate splitting $\Gamma-\Gamma_{\rm EP}\simeq\sqrt{\rho h}$ with
\begin{equation}
\rho=\frac{|K_M|}{T^2\zeta_0^2}=0.0158,\qquad
\Gamma_{\rm EP}=\frac{\ln|\zeta_0|}{T}.
\label{eq:sm_gap_growth_coefficient}
\end{equation}
The periodic modes are $\xi_\pm(k,t)=e^{-\Lambda_\pm t}U_k(t)v_\pm$ with $\Lambda_\pm=\Lambda_0\pm\delta\zeta/(T\zeta_0)$ to first order in $\delta\zeta$, and
\begin{equation}
\xi_\beta=\xi_0^{\rm per}+\beta\sqrt d\,\chi+O(d),\qquad
\chi=\frac{\delta\zeta}{\sqrt d}\xi_1^{\rm per},\qquad
\partial_k\xi_{\beta,a}=-\frac{\beta}{2\sqrt d}\chi_a+O(1),
\label{eq:sm_puiseux_modes}
\end{equation}
uniformly for $0\le t\le T$.
Away from nodes of $\xi_{0,a}^{\rm per}$,
\begin{equation}
\mathcal A_{\beta,a}=\beta\frac{\mathcal C_a(t)}{\sqrt d}+O(1),\qquad
\mathcal C_a=\frac12\frac{\Imm[\xi_{0,a}^{\rm per*}\chi_a]}{|\xi_{0,a}^{\rm per}|^2}.
\label{eq:sm_connection_coefficient}
\end{equation}

To extract the trembling coefficient from this expansion, we use the regular EP limits $f_a^{\rm EP}(t)$ of the sector fractions and the positive lower bound of $N_{\rm EP}=\sum_a c_a|\xi_{0,a}^{\rm per}|^2$ over one period.
With
\begin{equation}
\mathcal J_a^{(-1)}=\frac{c_a}{2}\Imm[\xi_{0,a}^{\rm per*}\chi_a],\qquad
\mathcal S_a=\frac{\mathcal J_a^{(-1)}-f_a^{\rm EP}\sum_j\mathcal J_j^{(-1)}}{N_{\rm EP}},\qquad
\mathcal L_{\beta,a}=\beta\frac{\mathcal S_a}{\sqrt d}+O(1),
\label{eq:sm_node_safe_ep_coefficient}
\end{equation}
where the last relation follows from Eq.~\eqref{eq:sm_node_safe_moments} uniformly, including nodes, we obtain
\begin{equation}
v_{{\rm src},\beta}=\beta\frac{\mathcal V(t)}{\sqrt d}+O(1),\qquad
\mathcal V=\frac{\dot\kappa}{\kappa}\mathcal S_P-2\eta\mathcal S_Q,\qquad
C_s=\left\langle(\mathcal V-\langle\mathcal V\rangle_T)^2\right\rangle_T^{1/2},
\label{eq:sm_source_ep_coefficient}
\end{equation}
which gives Eq.~\eqref{eq:trembling_scaling} whenever $C_s\ne0$, with $C_s=0.0822$.
Over one period the divergent part of $\partial_k x$ is $\mp(\partial_d\delta\zeta)\,b(t)$ for $x=U_k(t)v_\pm$, so the sector offsets arise from interference between the propagated chain vectors $a(t)$ and $b(t)$.

Integrating the same expansion over a fixed number of periods in the local narrow-band limit gives the mean-centered centroid displacement of Sec.~\ref{sec:sm_numerical_propagation},
\begin{equation}
\delta Z_\beta(t)=\frac{\beta}{\sqrt d}\bigl[\Upsilon(t)-\langle\Upsilon\rangle_I\bigr]+O(1),\qquad
\Upsilon(t)=\int_0^t\bigl[\mathcal V(t')-\langle\mathcal V\rangle_T\bigr]\,\mathrm{d}t',
\label{eq:sm_centered_displacement}
\end{equation}
with $\mathcal V$ of Eq.~\eqref{eq:sm_source_ep_coefficient}, $I$ the detrending interval of Sec.~\ref{sec:sm_numerical_propagation}, and the bounded flux contribution in the remainder.
A nonzero centered source waveform therefore gives the centered rms displacement the same exponent as $A_{\rm tr}$, with a coefficient that follows from the centered primitive and is not fixed by $C_s$ alone.

A two-band reduction illustrates how interference between the resonant harmonics enters $\mathcal C_a$.
Keeping the positive-frequency static branch in the $n=0$ harmonic and the negative-frequency branch in $n=-1$, with biorthogonal static eigenvectors $\ket{R_{s,\pm}}$, $\bra{L_{s,\pm}}$ and frequencies $\nu_\pm$, and projecting the modulation $\mathsf H_{\pm1}=-i(\kappa_0m/2)\,e_Qe_P^T$ of $\mathsf H=i\mathsf A$ onto them, $g=\bra{L_{s,+}}\mathsf H_1\ket{R_{s,-}}$, gives
\begin{equation}
\mathsf K=\begin{pmatrix}\Delta_{\rm red}&g\\-g^*&-\Delta_{\rm red}\end{pmatrix},\qquad
\omega_\beta=\omega_0+\beta\sqrt{\Delta_{\rm red}^2-|g|^2},
\label{eq:sm_two_band_reduction}
\end{equation}
with $\omega_0=(\nu_++\nu_-+\Omega)/2$ and $\Delta_{\rm red}=(\nu_+-\nu_--\Omega)/2$.
The form $\mathsf K_{21}=-\mathsf K_{12}^*$ holds in the conjugate-paired gauge $\ket{R_{s,-}}=\mathcal P\ket{R_{s,+}}^*$, $\bra{L_{s,-}}=\bra{L_{s,+}}^*\mathcal P$, which also gives $\nu_-=-\nu_+^*$ and real $\Delta_{\rm red}$.
In this reduction the EP lies at $k_{\rm EP}^{\rm red}=0.5691$ with $\rho_{\rm red}=-\partial_k(\Delta_{\rm red}^2-|g|^2)=0.0147$, distinct from the full values $k_{\rm EP}\simeq0.5744$ and $\rho=0.0158$.
With all EP quantities evaluated at $k_{\rm EP}^{\rm red}$ and the reduced detuning $d_{\rm red}=k_{\rm EP}^{\rm red}-k$, reconstructing the two harmonics gives the reduced coefficient
\begin{equation}
\mathcal C_a^{\rm red}(t)=
\frac{\sqrt{\rho_{\rm red}}}{2|g_{\rm EP}|}
\frac{\Imm\!\left[e^{-i\varphi_{\rm EP}}(R_{s,+;a}^{\rm EP})^*R_{s,-;a}^{\rm EP}e^{i\Omega t}\right]}
{\left|R_{s,+;a}^{\rm EP}+e^{-i\varphi_{\rm EP}}R_{s,-;a}^{\rm EP}e^{i\Omega t}\right|^2},
\label{eq:sm_reduced_connection}
\end{equation}
with $g_{\rm EP}=|g_{\rm EP}|e^{i\varphi_{\rm EP}}$.
Because the same-harmonic term is real, it drops out, so the $d^{-1/2}$ divergence comes from interference between the harmonics inherited from the positive- and negative-frequency static branches.

\section{Squared separations and the energy-norm Petermann factor}
\label{sec:sm_separations_petermann}
Each two-sector partition $A|B$ separates the spatial variance into internal sector variances $\sigma_A^2$ and $\sigma_B^2$ and a term set by the sector-centroid separation, $\sigma_z^2=f_A\sigma_A^2+f_B\sigma_B^2+f_Af_B(\Delta Z_{AB})^2$.
Modulation and damping select the partitions $P|\bar P$ and $Q|\bar Q$, which overlap, so $S_{\rm sep}=\sum_{a=P,Q}f_af_{\bar a}(\Delta Z_{a\bar a})^2$ is a source-selected moment distinct from $\sigma_z^2$.
Locally on the passband, $S_{\rm sep}=\sum_{a=P,Q}\mathcal L_{\beta,a}^2/(f_{\beta,a}f_{\beta,\bar a})$, with zero assigned to an empty sector or complement.
With $q=\beta\sqrt d$, the Cauchy--Schwarz inequality gives
\begin{equation}
0\le d\,\frac{\mathcal L_{\beta,a}^2}{f_{\beta,a}f_{\beta,\bar a}}
\le\frac{\sum_j c_j|\partial_q\xi_{\beta,j}|^2}{4N_\beta},
\label{eq:sm_separation_uniform_bound}
\end{equation}
a bound uniform over the period.
Together with the almost-everywhere limit
\begin{equation}
\begin{aligned}
d\,S_{\rm sep}\to c_{\rm sep}(t)
&=\sum_{a=P,Q}\frac{\mathcal S_a^2}{f_a^{\rm EP}f_{\bar a}^{\rm EP}}
=\sum_{a=P,Q}f_a^{\rm EP}f_{\bar a}^{\rm EP}(\mathcal C_a-\mathcal C_{\bar a})^2,\\
C_{\rm sep}&=\left\langle(c_{\rm sep}-\langle c_{\rm sep}\rangle_T)^2\right\rangle_T^{1/2},
\end{aligned}
\label{eq:sm_separation_rms_coefficient}
\end{equation}
where the second form holds away from nodes with $\mathcal C_{\bar a}=\sum_{j\ne a}f_j^{\rm EP}\mathcal C_j/f_{\bar a}^{\rm EP}$, it gives convergence in cycle rms, so $d\,A_{\rm sep}\to C_{\rm sep}$.
Its leading term is even in $\beta$, whereas the leading source velocity is odd.
For the medium of the main text $C_{\rm sep}=2.6$.

The energy Gramian defines the norms in the Petermann factor, whose divergence comes from the vanishing overlap of the left and right eigenvectors at the EP.
With $g_u(t)=\operatorname{diag}(\epsilon_b,\mu,\kappa/\alpha,1/\alpha)/4$, the period-averaged Gramian and, because the left eigenvector is a covector, the dual energy norm induced by $G_u^{-1}$ act as
\begin{equation}
\bra{R_\beta}G_u\ket{R_\beta}
=\frac1T\int_0^T\xi_\beta^\dagger g_u\xi_\beta\,\mathrm{d}t,\qquad
\bra{L_\beta}G_u^{-1}\ket{L_\beta}
=\frac1T\int_0^T\ell_\beta g_u^{-1}\ell_\beta^\dagger\,\mathrm{d}t,
\label{eq:sm_energy_dual_norms}
\end{equation}
since in the full harmonic space $G_u$ represents multiplication by $g_u(t)$ and $G_u^{-1}$ multiplication by $g_u(t)^{-1}$.
With $Mr_\beta=\zeta_\beta r_\beta$, $l_\beta M=\zeta_\beta l_\beta$ and $\Lambda_\beta=T^{-1}\ln\zeta_\beta$, the periodic right and left modes are $\xi_\beta(t)=e^{-\Lambda_\beta t}U_k(t)r_\beta$ and $\ell_\beta(t)=e^{\Lambda_\beta t}l_\beta U_k(t)^{-1}$, so $\ell_\beta(t)\xi_\beta(t)=l_\beta r_\beta$ and the denominator of Eq.~\eqref{eq:energy_petermann_factor} is $|l_\beta r_\beta|^2$.

Equation~\eqref{eq:sm_energy_dual_norms} gives the time-domain representation in the full harmonic space. For Fig.~\ref{fig:Fig2}(e), Eq.~\eqref{eq:energy_petermann_factor} is evaluated using the Sambe truncation specified in Sec.~\ref{sec:sm_model} and the inverse of the truncated energy Gramian. This inverse generally differs from the truncation of the
full inverse.
Rescaling either eigenvector leaves the factor $K_\beta^{(u)}\ge1$ of Eq.~\eqref{eq:energy_petermann_factor} unchanged.
Near the EP, $\ket{R_\beta}=\ket{R_0}+\beta\sqrt d\ket{R_1}$ and $\bra{L_\beta}=\bra{L_0}+\beta\sqrt d\bra{L_1}$ to first order in $\sqrt d$, with $\braket{L_0|R_0}=0$, so $\braket{L_\beta|R_\beta}=\beta\mathcal B\sqrt d+O(d)$ with $\mathcal B=\braket{L_1|R_0}+\braket{L_0|R_1}\ne0$, so
\begin{equation}
K_\beta^{(u)}=\frac{\mathcal K_0}{d}+O(d^{-1/2}),\qquad
\mathcal K_0=
\frac{\bra{R_0}G_u\ket{R_0}\bra{L_0}G_u^{-1}\ket{L_0}}{|\mathcal B|^2}>0,
\label{eq:sm_petermann_ep_coefficient}
\end{equation}
with $\mathcal K_0=0.0062$ here.
Consequently $A_{\rm tr}/\sqrt{K_\beta^{(u)}}\to C_s/\sqrt{\mathcal K_0}$ and $A_{\rm sep}/K_\beta^{(u)}\to C_{\rm sep}/\mathcal K_0$.

\section{Finite bandwidth and gap migration}
\label{sec:sm_finite_bandwidth}
For a finite packet, the source velocity and trembling amplitude are
\begin{equation}
\begin{aligned}
v_{{\rm src},\beta}^{\rm wp}
&=\frac{\dot\kappa}{\kappa}f_{\beta,P}^{\rm wp}f_{\beta,\bar P}^{\rm wp}\Delta Z_{\beta,P\bar P}^{\rm wp}
-2\eta f_{\beta,Q}^{\rm wp}f_{\beta,\bar Q}^{\rm wp}\Delta Z_{\beta,Q\bar Q}^{\rm wp},\\
A_{\rm tr}^{\rm wp}
&=\left\langle\left(v_{{\rm src},\beta}^{\rm wp}-\langle v_{{\rm src},\beta}^{\rm wp}\rangle_T\right)^2\right\rangle_T^{1/2},
\end{aligned}
\label{eq:sm_packet_source_trembling}
\end{equation}
where the mean and rms are taken over one period, the first period unless stated otherwise.
In the joint limit $d_0\to0^+$, $\sigma_k/d_0\to0$ over a fixed window and away from nodes, a passband packet has
\begin{equation}
\Delta Z_{\beta,a\bar a}^{\rm wp}
=\beta\frac{\mathcal C_a-\mathcal C_{\bar a}}{\sqrt{d_0}}+o(d_0^{-1/2}),
\label{eq:sm_joint_narrowband_separation}
\end{equation}
an additive error that also applies where $\mathcal C_a=\mathcal C_{\bar a}$.

For the Gaussian average, take a regular launch with the phase convention of Sec.~\ref{sec:sm_packets_gap_null} and the Gaussian weight $w(k)\propto\exp[-(k-k_0)^2/(2\sigma_k^2)]$, and write the packet moments with the normalized spectral-energy measure $\mathrm{d}m_t\propto w(k)N_\beta(k,t)e^{2\Gamma_\beta(k)t}\,\mathrm{d}k$, with $N_\beta$ of Eq.~\eqref{eq:sm_node_safe_moments}.
At fixed $\lambda$, over the first period and within a few $\sigma_k$ of $k_{\rm EP}$, the relative growth factor $e^{2(\Gamma_\beta-\Gamma_{\rm EP})t}$, the sector weights $n_{\beta,a}$ and their sum $N_\beta$ differ from $1$, $n_a^{\rm EP}$ and $N_{\rm EP}>0$ by $O(\sqrt{\sigma_k})$ on both sides of the EP, so the absolute expansions of the sector weights stay valid at nodes.
The weighted offset $f_a^{\rm wp}f_{\bar a}^{\rm wp}\Delta Z_{a\bar a}^{\rm wp}$ equals the $m_t$-average of the local offset $\mathcal L_{\beta,a}$ of Eq.~\eqref{eq:sm_node_safe_moments} plus the $m_t$-covariance of $f_a$ with the connection $B_z=\mathcal A_{\rm env}+t\,\partial_k\omega_{\beta,R}+\mathcal A_u$, where $\mathcal A_u=\sum_j\mathcal J_{\beta,j}/N_\beta$ is the energy-weighted connection of the full mode.
Although $B_z$ is singular at the EP, $f_a-f_a^{\rm EP}=O(|k-k_{\rm EP}|^{1/2})$ while the singular part of $B_z$ is $O(|k-k_{\rm EP}|^{-1/2})$, so the covariance is $O(1)$ and subleading to the $\sigma_k^{-1/2}$ waveform.
Gaussian tails outside a fixed $k$-neighborhood of the EP are exponentially small as $\sigma_k\to0$ at fixed $\lambda$.
With no local source from the gap side (Sec.~\ref{sec:sm_packets_gap_null}) and the common waveform $v_{\rm tr}(k,t)=d^{-1/2}w_{\rm tr}(t)$ on the passband, the packet waveform is
\begin{equation}
\begin{aligned}
v_{\rm tr}^{\rm wp}(t)
&=w_{\rm tr}(t)\int_0^\infty\frac{\mathrm{d}d}{\sqrt{2\pi}\,\sigma_k}
\frac{e^{-(d-d_0)^2/(2\sigma_k^2)}}{\sqrt d}+o(\sigma_k^{-1/2})\\
&=w_{\rm tr}(t)\,\sigma_k^{-1/2}F(\lambda)+o(\sigma_k^{-1/2})
\end{aligned}
\label{eq:sm_gaussian_waveform}
\end{equation}
in cycle rms, which gives Eq.~\eqref{eq:finite_bandwidth_scaling}.
Substituting $u=v^2$ for the integration variable $u=d/\sigma_k$ of Eq.~\eqref{eq:finite_bandwidth_scaling} removes the endpoint singularity.
$F(0)=2^{-3/4}\Gamma(1/4)/\sqrt{2\pi}=0.860$, $F$ peaks at $1.02$ for $\lambda=0.77$, $F=\lambda^{-1/2}[1+3/(8\lambda^2)+O(\lambda^{-4})]$ for $\lambda\gg1$, and $F\simeq e^{-\lambda^2/2}/\sqrt{2|\lambda|}$ for $\lambda\to-\infty$.
The average holds while $d_0$ and $\sigma_k$ lie inside the square-root region: for $\sigma_k=10^{-5}$ and $10^{-4}$ in Fig.~\ref{fig:Fig3}(a) the first-period amplitude departs from $C_sF(\lambda)/\sqrt{\sigma_k}$ by about $5\%$ at $d_0\simeq5\times10^{-4}$ and $10\%$ at $d_0\simeq10^{-3}$, and for $\sigma_k=10^{-3}$ a finite-bandwidth correction is already present near $d_0=0$.

At later times relative amplification moves spectral weight into the gap, where the branch with the larger growth rate has $\Gamma(h)=\Gamma_{\rm EP}+\sqrt{\rho h}+O(h)$ with $h=k-k_{\rm EP}$ and $\rho$ of Eq.~\eqref{eq:sm_gap_growth_coefficient}.
The common factor $e^{2\Gamma_{\rm EP}t}$ cancels from the normalized spectrum, and with $y=h/\sigma_k$ and $\tau=t\sqrt{\rho\sigma_k}$ the leading exponent is
\begin{equation}
\Phi_+(y;\lambda,\tau)=-\tfrac12(y+\lambda)^2+2\tau\sqrt y,
\label{eq:sm_gap_spectral_exponent}
\end{equation}
whose interior stationary point gives Eq.~\eqref{eq:spectral_saddle} and $h_s=\sigma_ky_s$.
On the opposite gap branch, however, the square-root term changes sign, so for $\lambda\ge0$ no interior saddle forms.
The saddle describes the spectral maximum once its contribution dominates the passband weight, $h_s$ lies inside the gap-side square-root region, and the $O(h)$ remainder of $\Gamma$ times $t$ stays small, so the relevant spectral distance from the EP is $d_0$ for a resolved passband packet, $\sigma_k$ for a packet overlapping the EP and $h_s$ once the gap saddle dominates.
For this medium $\Gamma_{\rm EP}=-2.6\times10^{-3}\Omega$, so relative growth reshapes the normalized spectrum while the absolute energy decays.
The leading local estimate for the onset of net growth on the gap side is $h\simeq\Gamma_{\rm EP}^2/\rho\approx4\times10^{-4}$, and the energy minimum of Fig.~\ref{fig:Fig3}(c) occurs near the stage when the saddle reaches this scale.

\section{Numerical propagation}
\label{sec:sm_numerical_propagation}

We use fourth-order Runge--Kutta integration to propagate $x=(E,H,P,Q)^T$ and its spectral derivative on the transformed grid $k=k_{\rm EP}+\vartheta|\vartheta|$.
The derivative $\partial_\vartheta x=(\mathrm{d}k/\mathrm{d}\vartheta)\partial_kx$ obeys
\begin{equation}
\begin{aligned}
\partial_t(\partial_\vartheta x)
&=\mathsf A(k,t)\,\partial_\vartheta x
+\frac{\mathrm{d}k}{\mathrm{d}\vartheta}
 \partial_k\mathsf A(k,t)x,\\
\frac{\mathrm{d}k}{\mathrm{d}\vartheta}&=2|\vartheta|.
\end{aligned}
\label{eq:sm_augmented_propagation}
\end{equation}
Sector energies and centroids are evaluated using Eq.~\eqref{eq:sm_parseval_centroid}, with trapezoidal quadrature in $\vartheta$ for Fig.~\ref{fig:Fig2}(d) and midpoint quadrature for Fig.~\ref{fig:Fig3}.
In the latter, the EP is a cell boundary and is not sampled directly.
Initial spectral derivatives are obtained by finite differences for Fig.~\ref{fig:Fig2}(d) and by differentiating the monodromy eigenproblem for Fig.~\ref{fig:Fig3}. Long-time evolution is evaluated from the one-period Floquet data.

Each eigen-packet uses a real Gaussian envelope with spectral-intensity rms width $\sigma_k$, multiplying $\xi_\beta(k,0)$ normalized to unit instantaneous energy with its electric component real and positive.
A real envelope contributes only its weight to the energy and first-moment integrals.
On the gap side, the $\beta=-$ passband branch is continued
into the branch with the larger growth rate.

The medium has $\epsilon_b=\mu=\Omega=1$, $\alpha=0.6$, $\kappa_0=2$, $m=0.25$, and $\eta=0.08$, with $k_{\rm EP}=0.5743977113$ determined from the discriminant of the resonant multiplier pair.
The trajectories of Fig.~\ref{fig:Fig2}(d) use $\sigma_k=d_0/20$, $801$ nodes over $k_0\pm6\sigma_k$, and a time step $T/800$ over four periods.
With $Z_{\rm det}(t)=Z(t)-\bar vt$, $\bar v=[Z(4T)-Z(T)]/(3T)$, and $I=[T,4T]$,
the plotted displacement is $\delta Z=Z_{\rm det}-\langle Z_{\rm det}\rangle_I$.
The flux contribution $\int_0^t v_{\rm flux}(t')\,\mathrm{d}t'$ is detrended and centered over the same interval.
The calculations of Fig.~\ref{fig:Fig3} use a time step $T/12800$ for the one-period propagation; the long-time packet in panels (b,c) is integrated over $k_{\rm EP}-9\sigma_k\le k\le k_{\rm EP}+16\sigma_k$.

The coefficients of Fig.~\ref{fig:Fig_loss} come from the local source velocity, the band slope and the time-domain Petermann factor at $d=10^{-4}$, $10^{-5}$ and $10^{-6}$ from the EP of each loss profile, with a time step $T/2000$, extrapolated to $d\to0$ by a quadratic fit in $d$.
The field-only packets of Fig.~\ref{fig:Fig_field} use $801$ nodes over $k_0\pm6\sigma_k$, the one-period monodromy from a time step $T/2000$, and the centered rms of $v_{\rm src}$, sampled within each period at $T/400$ and displayed every third period in Fig.~\ref{fig:Fig_field}(a); the displacement of Fig.~\ref{fig:Fig_field}(d) is detrended by a least-squares straight line over that period, which leaves a zero-mean residual.
The monochromatic calculation uses the monodromy at a single wavenumber from a time step $T/4000$. For both monochromatic and packet launches, $A_b$ is the largest one-period centered rms of $v_{\rm src}$ over periods $I_n=[nT,(n+1)T]$ whose midpoints lie in $[0.7t_b,1.3t_b]$. The packet calculations use $801$ nodes at $d_0=10^{-3}$ and $1601$ nodes at smaller $d_0$, monodromies from a time step $T/2000$ ($T/4000$ at $d_0=10^{-6}$), and sampling at $T/400$ or $T/800$ within the period.
The late-time regularized comparison of Sec.~\ref{sec:sm_finite_variance} uses $6000$ nodes on the transformed grid over $k_{\rm EP}-9\sigma_k\le k\le k_{\rm EP}+16\sigma_k$ with the monodromy from a time step $T/4000$.

\section{Medium dependence}
\label{sec:sm_medium_dependence}
Modulation and damping determine $C_s$ through the rms amplitudes and correlation of their leading source waveforms.
With $w_{P,Q}(t)=\lim_{d\to0^+}\sqrt d\,[v_{P,Q}-\langle v_{P,Q}\rangle_T]$, their rms values $C_P$ and $C_Q$ over one period and, when both are nonzero, their correlation coefficient $\varrho$, the total coefficient obeys
\begin{equation}
C_s^2=C_P^2+C_Q^2+2\varrho\,C_PC_Q.
\label{eq:sm_channel_correlation}
\end{equation}

Without damping the modulation channel alone acts and the chain coefficients of Eq.~\eqref{eq:sm_source_ep_coefficient} give $C_P\simeq0.3\,m$ for modulation depths between $0.1$ and $0.4$.
Damping acts through the kinetic-energy sector and, in the media we computed, raises $C_Q$ faster than it changes $C_P$ and adds the two channels with positive correlation, so that at $m=0.25$ and $\eta=0.5$ the damping channel ($C_Q=0.093$) exceeds the modulation channel ($C_P=0.058$).
For $m=0.25$, a modulated loss rate $\eta(t)=\eta_0[1+\delta\cos(\Omega t+\varphi)]$ with $0\le\delta\le1$ enters through the source term $-2\eta(t)u_Q$ and the Floquet modes, and each loss profile has its own EP.
At each EP we compare the trembling coefficient $C_s$ with the band-slope coefficient and the Petermann coefficient,
\begin{equation}
C_g=\lim_{d\to0^+}\sqrt d\,|\partial_k\Ree\omega_-|,\qquad
\mathcal K_0=\lim_{d\to0^+}d\,K_-^{(u)},
\label{eq:sm_loss_coefficients}
\end{equation}
with $C_s$ from the local source velocity of Eq.~\eqref{eq:sm_local_source} and $K_-^{(u)}$ from the time-domain representation of Eq.~\eqref{eq:sm_energy_dual_norms}.
Measured from the restoring coefficient, $\varphi=0$ and $\pi$ give loss in phase and in antiphase with $\kappa(t)$, as can occur at linear order when a common control modulates both coefficients with negligible relative response lag, and $\varphi=\pi/2$ and $3\pi/2$ give loss in quadrature.
For the medium of Fig.~\ref{fig:Fig2}, $\eta_0=0.08$, in-phase and antiphase loss change $C_s$ monotonically with $\delta$, lowering it by about $6\%$ and raising it by about $11\%$ at $\delta=1$, while $C_g$ and $\mathcal K_0$ barely change [Fig.~\ref{fig:Fig_loss}(a)].
Quadrature loss at this damping moves all three coefficients by comparable fractions.
Single-branch packets centered $d_0=10^{-4}$ below each profile's EP, with $\sigma_k=d_0/20$, have first-period trembling amplitudes $11\%$ higher for full-depth antiphase loss and $6\%$ lower for in-phase loss than under constant loss.
Their centroid displacements, detrended over $[T,4T]$ as in Sec.~\ref{sec:sm_numerical_propagation}, change in centered rms by $+10\%$ and $-5\%$ for $Z$, respectively, and by less than $1\%$ for $Z_{\rm EM}$.
At $\eta_0=0.3$ and $\delta=1$ the three coefficients follow different curves as the loss phase varies [Fig.~\ref{fig:Fig_loss}(b)].
Near $\varphi=\pi$ the trembling coefficient exceeds its constant-loss value by about a quarter while $C_g$ and $\mathcal K_0$ return close to their constant-loss values, and near $\varphi=3\pi/2$ the trembling coefficient returns close to its constant-loss value while $C_g$ and $\mathcal K_0$ fall by about a fifth and a third.

Over the same media, with $\eta$ up to $0.5$ at $m=0.25$ and up to $0.3$ at $m=0.1$ and $0.4$, the single-branch amplitude $A_{\rm tr}(d)$ follows the inverse-square-root scaling of Eq.~\eqref{eq:trembling_scaling} between $d=10^{-5}$ and $10^{-4}$.

\section{Finite-variance launch}
\label{sec:sm_finite_variance}
We implement the regularization of Sec.~\ref{sec:sm_packets_gap_null} by replacing the launch profile $\xi_-(k,0)$ on $|k-k_{\rm EP}|<\varepsilon_k/2$ with the linear interpolation between its values at the two ends of that interval, and we propagate the smoothed packets with the full Maxwell--Lorentz equations.
This interpolant is continuous and adds no rapid phase.
Inside the interval, however, it generally does not keep the gap-side reality of Sec.~\ref{sec:sm_packets_gap_null}, which holds outside it, and this local change enters the moments at the order estimated there.

\begin{figure*}[tbp]
\centering
\includegraphics[width=\textwidth]{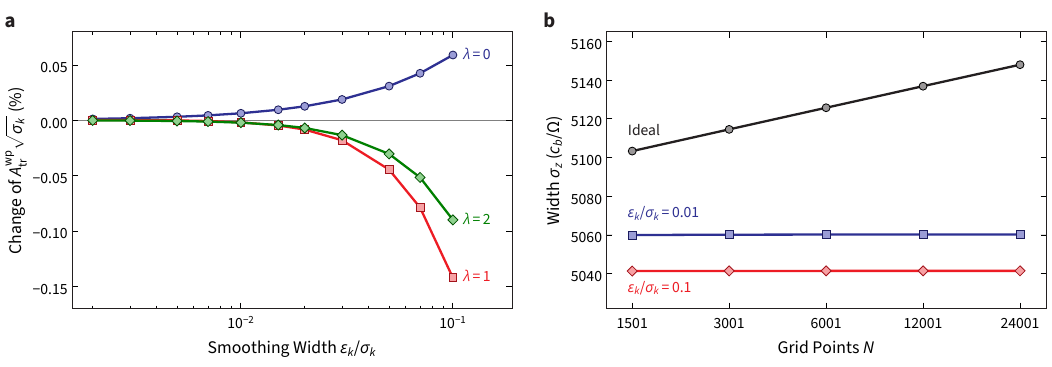}
\caption{\label{fig:S_variance}
Finite-variance launch with $\sigma_k=10^{-4}$.
(a) Relative change of $A_{\rm tr}^{\rm wp}\sqrt{\sigma_k}$ over the first period against the smoothing width $\varepsilon_k/\sigma_k$, for $\lambda=0$, $1$ and $2$.
(b) Spatial rms width $\sigma_z$ at launch against the number $N_k$ of points on the grid $k=k_{\rm EP}+\vartheta|\vartheta|$, for the ideal launch and for $\varepsilon_k/\sigma_k=0.01$ and $0.1$ ($\lambda=0$). Symbols mark computed points.}
\end{figure*}

For $\sigma_k=10^{-4}$ and $\varepsilon_k\le0.1\,\sigma_k$ the scaled amplitude $A_{\rm tr}^{\rm wp}\sqrt{\sigma_k}$ changes by less than $0.15\%$ [Fig.~\ref{fig:S_variance}(a)], and the smoothed packets keep $\max_t|v_{\rm flux}|\le0.17\,c_b$.
For the ideal launch, $\sigma_z^2$ grows with the logarithm of the number of $k$ points as the integral $\int|\partial_kx|^2\,\mathrm{d}k$ of Sec.~\ref{sec:sm_packets_gap_null} is resolved, whereas the smoothed launches have a fixed width $\sigma_z\simeq5.0\times10^3\,c_b/\Omega$ [Fig.~\ref{fig:S_variance}(b)].
For $\sigma_k=10^{-4}$ and $\lambda=2$, the smoothed launches with $\varepsilon_k=0.01\,\sigma_k$ and $0.1\,\sigma_k$, propagated with powers of the monodromy as in Sec.~\ref{sec:sm_chirp}, follow the decay of the ideal launch through $\tau=10$, where all three scaled amplitudes are of order $10^{-9}$.

\section{Field-energy and electric-energy centroids}
\label{sec:sm_observables}
\begin{figure*}[tbp]
\centering
\includegraphics[width=\textwidth]{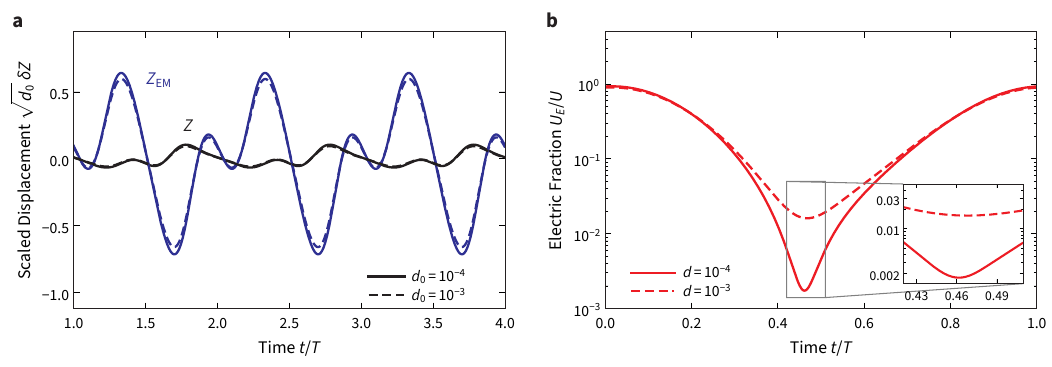}
\caption{\label{fig:S_observables}
(a) Detrended, centered displacements $\delta Z$ of the total-energy centroid $Z$ and the field-energy centroid $Z_{\rm EM}$, multiplied by $\sqrt{d_0}$ (units $(c_b/\Omega)^{1/2}$), for single-branch packets with $\sigma_k=d_0/20$ at $d_0=10^{-3}$ and $10^{-4}$.
(b) Electric-energy fraction $U_E/U$ of the single-branch mode over one period at $d=10^{-3}$ and $10^{-4}$. The inset enlarges the dip.}
\end{figure*}

Measurements of $E$ and $H$ give access to the centroid of the field energy,
\begin{equation}
Z_{\rm EM}=\frac{\int z\,(u_E+u_H)\,\mathrm{d}z}{\int(u_E+u_H)\,\mathrm{d}z},
\label{eq:sm_field_centroid}
\end{equation}
and of the electric energy alone, $Z_E$.
For the packets of Fig.~\ref{fig:Fig2}(d), detrended and centered as in Sec.~\ref{sec:sm_numerical_propagation}, $Z_{\rm EM}$ has a centered rms displacement about eight times that of $Z$, with a different waveform, and multiplying both displacements by $\sqrt{d_0}$ makes the two detunings nearly coincide [Fig.~\ref{fig:S_observables}(a)].
The flux contribution to the displacement has an rms of at most a few percent of the total.
The electric fraction $U_E/U$ of the single-branch mode dips within each cycle, and the dip deepens as the EP is approached [Fig.~\ref{fig:S_observables}(b)].
Near the dip the small denominator makes $Z_E$ sensitive to the electric-sector first moment, and between the two evaluated detunings its displacement increases more strongly than those of $Z$ and $Z_{\rm EM}$.

\section{Spectral chirp}
\label{sec:sm_chirp}
Multiplying the launch by $e^{i\phi_{\rm ch}(k)}$ replaces $i\partial_k\mathcal F_a$ by $e^{i\phi_{\rm ch}}(i\partial_k\mathcal F_a-\partial_k\phi_{\rm ch}\,\mathcal F_a)$ in Eq.~\eqref{eq:sm_parseval_centroid}, so each sector centroid shifts as
\begin{equation}
Z_a\to Z_a-\langle\partial_k\phi_{\rm ch}\rangle_a,\qquad
\langle\partial_k\phi_{\rm ch}\rangle_a=\frac{\int\partial_k\phi_{\rm ch}\,|\mathcal F_a|^2\,\mathrm{d}k}{\int|\mathcal F_a|^2\,\mathrm{d}k},
\label{eq:sm_chirp_shift}
\end{equation}
and its offset from the total centroid changes by $-\langle\partial_k\phi_{\rm ch}\rangle_a+\langle\partial_k\phi_{\rm ch}\rangle_{\rm tot}$.
A constant phase leaves every centroid unchanged and a linear phase shifts all of them equally.
A nonlinear phase, however, can give unequal shifts when the sector weights depend differently on $k$, and can lift the gap-side null of Sec.~\ref{sec:sm_packets_gap_null}.

At fixed $U$ and $f_a$ the source velocity is linear in the sector first moments $J_a=U_aZ_a$,
\begin{equation}
v_{\rm src}=\mathcal W_t[J]\equiv\frac1U\Bigl[\frac{\dot\kappa}{\kappa}\Bigl(J_P-f_P\sum_bJ_b\Bigr)-2\eta\Bigl(J_Q-f_Q\sum_bJ_b\Bigr)\Bigr],
\label{eq:sm_linear_source}
\end{equation}
so any split of the $J_a$ into parts gives source waveforms that add, while their centered rms values do not.
Restricting the $k$ integrals of $J_a$ to $k>k_{\rm EP}$ in Eq.~\eqref{eq:sm_linear_source} gives the gap-side part.

For the quadratic chirp $\phi_{\rm ch}=(k-k_0)^2/(4\sigma_k^2)$ with $\sigma_k=10^{-4}$ and $\lambda=2$, propagated with powers of the monodromy, the gap-side first-moment contribution approaches the full chirped response at the later computed times, and that response decreases slowly up to $\tau=20$ [Fig.~\ref{fig:S_preparation}(a)].
Without the chirp, the same launch fades [Fig.~\ref{fig:S_preparation}(a)].

\section{Launch in the fields alone}
\label{sec:sm_field_launch}
\begin{figure*}[tbp]
\centering
\includegraphics[width=0.62\textwidth]{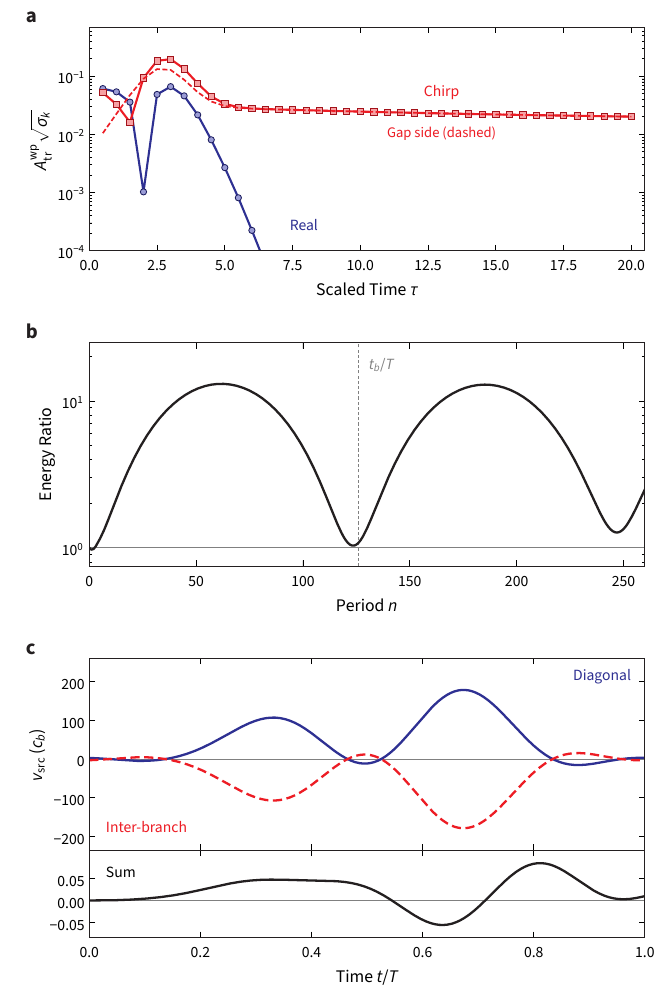}
\caption{\label{fig:S_preparation}
Preparation dependence of the trembling.
(a) $A_{\rm tr}^{\rm wp}\sqrt{\sigma_k}$ over the period nearest each $\tau$ for the real-envelope launch and for the quadratic chirp, with the gap-side part of the chirped response, for $\sigma_k=10^{-4}$ and $\lambda=2$. The plotted range ends at $A_{\rm tr}^{\rm wp}\sqrt{\sigma_k}=10^{-4}$, and the real-envelope response continues below it.
(b) Energy of the field-only packet divided by that of a single-branch packet with the same envelope, both normalized at launch, for $d_0=10^{-3}$ and $\sigma_k=d_0/20$; vertical dashed line, $t_b/T$ of Eq.~\eqref{eq:sm_beat_time}.
(c) Branch-diagonal and inter-branch source waveforms of the field-only packet over the first period at $d_0=10^{-4}$ (upper axis) and their sum on an expanded scale (lower axis).
Parameters as in Fig.~\ref{fig:Fig2}.}
\end{figure*}

The launch $x_0=(1,1,0,0)^{\rm T}$ is a forward field pulse with the polarization at rest.
Expanding it in the right Floquet eigenvectors $r_j$ of the monodromy defines the diagonal energies
\begin{equation}
x_0=\sum_jc_jr_j,\qquad
e_j=\frac{|c_j|^2\,r_j^\dagger g_u(0)\,r_j}{x_0^\dagger g_u(0)\,x_0},
\label{eq:sm_modal_energies}
\end{equation}
and the interference energy $1-\sum_je_j$.
The diagonal energies of the resonant pair grow as $1/d$ and reach about $30$ at $d=10^{-4}$, and the negative interference energy cancels the excess in their sum, returning the normalized total to one.
Splitting $J_a$ into branch-diagonal and inter-branch parts in Eq.~\eqref{eq:sm_linear_source} gives two source waveforms of nearly equal and opposite shape, which cancel to a small remainder [Fig.~\ref{fig:S_preparation}(c)].

The two critical branches split as $\Ree\omega_\pm\simeq\Ree\omega_{\rm EP}\pm2C_g\sqrt d$, so their relative phase returns after
\begin{equation}
t_b=\frac{2\pi}{|\Ree\omega_+-\Ree\omega_-|}\simeq\frac{\pi}{2C_g\sqrt d},
\label{eq:sm_beat_time}
\end{equation}
while the relative group velocity $2C_g/\sqrt d$ separates the two branch components by about $\pi/d$.
Near half a beat the resonant components add, and to leading order the energy relative to a single-branch packet, both normalized at launch, approaches $(\sqrt{e_+}+\sqrt{e_-})^2\simeq0.013/d$ [Fig.~\ref{fig:S_preparation}(b)].
Near $t_b$ the energy returns to its interference minimum, the components no longer cancel in the first moments, and the trembling bursts [Fig.~\ref{fig:Fig_field}(a)].

In the monochromatic limit we propagate $x_0$ at a single wavenumber with powers of the monodromy and evaluate the burst amplitude using the common maximum-rms definition of Sec.~S7. The computed burst amplitudes follow
\begin{equation}
A_b\simeq C_bd^{-3/2},\qquad C_b\simeq0.0031,
\label{eq:sm_burst_law}
\end{equation}
over $10^{-6}\le d\le10^{-3}$ [Fig.~\ref{fig:Fig_field}(b)].
We write the state as a sum of the critical pair and a noncritical part.
The noncritical multipliers have modulus $0.79$ per period against $0.98$ for the critical pair, so their contribution is negligible at $t_b$.
The critical amplitudes, of order $d^{-1/2}$ and opposite sign, cancel at launch and again near the recurrence, where the relative phase returns, and the decay-removed energy at the recurrence minimum stays between $0.95$ and $0.98$ of the launch energy.
The wavenumber derivative of the accumulated relative phase reaches
\begin{equation}
t_b\,|\partial_k(\Ree\omega_+-\Ree\omega_-)|\simeq\frac{\pi}{d}
\label{eq:sm_phase_derivative}
\end{equation}
at fixed observation time near $t_b$.
Multiplied by branch amplitudes of order $d^{-1/2}$, this phase gradient therefore permits a $d^{-3/2}$ contribution to the first moments and the source velocity when the decay-removed recurrence state remains finite.

For $d_0=10^{-3}$ [Fig.~\ref{fig:Fig_field}(c,d)] the burst peaks about $123$ periods after launch and lasts about $20$ periods, when the pulse retains about $2\%$ of its launch energy, close to the single-branch packet at the same time, and the detrended centroid swings by about $150\,c_b/\Omega$ peak to peak within the burst period, with a largest excursion of about $90\,c_b/\Omega$, near $1\%$ of the initial rms width $1/(2\sigma_k)=10^4\,c_b/\Omega$.
For the monochromatic recurrence, whose decay-removed energy stays of order one, the absolute energy at the recurrence is about $e^{2\Gamma_{\rm EP}t_b}$, which falls from $2\times10^{-2}$ at $d=10^{-3}$ to $2\times10^{-6}$ at $10^{-4}$.

For a packet the beat phase varies across the spectrum by $|\partial_k(\Ree\omega_+-\Ree\omega_-)|\,t_b\,\sigma_k=\pi\sigma_k/d_0$ at $t_b$.
To leading order, neglecting the monochromatic baseline offset of the launch-normalized energy ratio from one, this raises the energy ratio at the recurrence minimum above one by
\begin{equation}
\Delta R\simeq\Bigl(\frac{\pi\sigma_k}{2d_0}\Bigr)^2\bigl(\sqrt{e_+}+\sqrt{e_-}\bigr)^2\simeq0.031\,\frac{\sigma_k^2}{d_0^3},
\label{eq:sm_refilling}
\end{equation}
which the computed minima approach as $d_0$ decreases [Fig.~\ref{fig:Fig_field}(b), inset].
Approaching the monochromatic burst thus requires $\Delta R\ll1$, and the packet bursts fall increasingly below the monochromatic values as this excess becomes appreciable [Fig.~\ref{fig:Fig_field}(b)].
A launch that is real in the frame $(E,-iH,P,Q)$, such as equal forward and backward field pulses with a real envelope and zero relative phase, stays real and gives no source velocity at any time.

\end{document}